\documentclass[aps,prd,noshowpacs,preprintnumbers,nofootinbib,twoside,twocolumn]{revtex4}

\usepackage{amsmath}
\usepackage{amssymb}
\usepackage{fancyhdr}
\usepackage{bm}

\newcommand{\Cal}[1]{\ensuremath{\mathcal{#1}}}

\newcommand{\eqn}[1]{Eqn. \eqref{#1}}
\newcommand{\eqns}[1]{Eqns. \eqref{#1}}
\newcommand{\Cite}[1]{Ref. \cite{#1}}
\newcommand{\Cites}[1]{Refs. \cite{#1}}
\newcommand{\ph}[1]{\phantom{#1}}

\newcommand{\avg}[1]{\ensuremath{\left\langle #1\right\rangle}}
\newcommand{\stavg}[1]{\ensuremath{\left\langle #1\right\rangle_{ST}}}
\newcommand{\avgD}[1]{\ensuremath{\left\langle #1\right\rangle_{\Cal{D}}}}

\newcommand{\om}{{\bm{\omega}}} 
\newcommand{\Om}[2]{{\bm{\Omega}{}^{#1}_{\ph{#1}#2}}} 
\newcommand{\barOm}[2]{\bar{\bm{\Omega}}{}^{#1}_{\ph{#1}#2}} 
\newcommand{\rmb}[1]{{\bf #1}} 
\newcommand{\bZ}[4]{{\rmb{Z}{}^{#1\ph{#2}#3}_{\ph{#1}#2\ph{#3}#4}}}
\newcommand{\ext}{{\rmb{d}}} 
\newcommand{\dx}[1]{\ext x^{#1}} 

\newcommand{\dbil}{{\ext\negthickspace\negmedspace^{-}}}

\newcommand{\Dbil}[1]{{\rmb{D}_{#1}\negthickspace\negthickspace
    \negthickspace\negthickspace\negthickspace\negthinspace
    \text{--\thickspace\thickspace\thickspace\,\,\,}}}

\newcommand{\bil}[1]{\widetilde{#1}}

\newcommand{\Wxx}[2]{\ensuremath{\Cal{W}{}^{#1^\prime}_{#2}(x^\prime,x)}}
\newcommand{\Wxxinv}[2]{\ensuremath{\Cal{W}{}^{#1}_{#2^\prime}(x,x^\prime)}}
\newcommand{\W}[2]{\ensuremath{\Cal{W}{}^{#1^\prime}_{#2}}}

\newcommand{\Linh}{\ensuremath{L_{\rm inhom}}}
\newcommand{\Lfrw}{\ensuremath{L_{\rm FLRW}}}
\newcommand{\Lhub}{\ensuremath{L_{\rm Hubble}}}
\newcommand{\bt}{\ensuremath{\bar t}}
\newcommand{\ba}{\ensuremath{\bar a}}
\newcommand{\Mbar}{\ensuremath{\bar{\Cal{M}}}}

\newcommand{\calD}{\ensuremath{\mathcal{D}}}
\newcommand{\aD}{\ensuremath{a_\calD}}

\newcommand{\uD}[1]{\ensuremath{#1_\Cal{D}}}

\begin{document}

\title{The Spatial Averaging Limit of Covariant Macroscopic Gravity\break
{- {\it Scalar\ Corrections\ to\ the\ Cosmological\ Equations} -}}

\author{Aseem Paranjape}
\email{aseem@tifr.res.in}

\author{T. P. Singh}
\email{tpsingh@tifr.res.in}

\affiliation{Tata Institute of Fundamental Research, Homi Bhabha
Road, Colaba, Mumbai - 400 005, India\\}


\begin{abstract}
\noindent It is known that any explicit averaging scheme of the
type essential for describing the large scale behaviour of the
Universe, must necessarily yield corrections to the Einstein
equations applied in the Cosmological setting. The question of
whether or not the resulting corrections to the Einstein equations
are significant, is still a subject of debate, partly due to
possible ambiguities in the averaging schemes available. In
particular, it has been argued in the literature that the effects
of averaging could be gauge artifacts.  We apply the formalism of
Zalaletdinov's Macroscopic Gravity (MG) which is a fully covariant
and nonperturbative averaging scheme, in an attempt to construct
gauge independent corrections to the standard
Friedmann-Lema\^itre-Robertson-Walker (FLRW) equations. We find
that whereas one cannot escape the problem of dependence on
\emph{one} gauge choice -- which is inherent in the assumption of
large scale homogeneity and isotropy -- it is however possible to
construct \emph{spacetime scalar} corrections to the standard FLRW
equations. This partially removes the criticism concerning the
corrections being gauge artifacts. For a particular initial choice
of gauge which simplifies the formalism, we explicitly construct
these scalars in terms of the underlying inhomogeneous geometry,
and incidentally demonstrate that the formal structure of the
corrections with this gauge choice is identical to that of
analogous corrections derived by Buchert in the context of spatial
averaging of scalars.
\end{abstract}

\maketitle

\section{Introduction}
\noindent One of the basic assumptions of Cosmology is that the
matter distribution in the Universe, when averaged on large enough
scales, is homogeneous and isotropic. One then models the metric
of spacetime on such large scales as having the
Friedmann-Lema\^itre-Robertson-Walker (FLRW) form, i.e. -- having
homogeneous and isotropic spatial sections, and applies Einstein's
General Relativistic field equations to determine the geometry of
the Universe. The idea here is that just like the matter, the
metric can also be ``averaged'' on large scales to yield a highly
symmetric geometry. The averaging operation is usually an implicit
(and often vague) assumption. It has long been known however
\cite{ellis}, that any \emph{explicit} averaging scheme for the
metric of spacetime and energy-momentum tensor of matter, must
necessarily yield corrections to the Einstein equations -- these
equations being ideally imposed on length scales where stars (and
not galaxies) are pointlike objects.

The problem of constructing an explicit and self-consistent
averaging scheme for the Einstein equations in the Cosmological
setting has a long history \cite{averaging}. Recently, attention
has been devoted to an averaging scheme developed by Buchert
\cite{buchDust,buchPress}, who defines a spatial averaging
operation for scalar quantities within a chosen $3+1$ splitting of
spacetime, and derives effective equations which could have
nontrivial effects on the assumed, homogeneous and isotropic model
of the Universe. The question of whether or not the resulting
corrections to the Einstein equations are significant, is still a
subject of debate \cite{kolb,chung,wald,rasanen} and can be
considered to be open. Buchert's approach to averaging has also
been debated on the grounds that it is concerned with spatial
averages, and not spacetime averages.

In such a context, it is perhaps desirable to work with an
averaging scheme which starts from a spacetime averaging, and
which allows us to construct the corrections to Einstein's
equations as physically relevant, gauge independent objects that
can (in principle at least) be observationally tested. An
averaging scheme for tensors on an arbitrary pseudo-Riemannian
manifold has been developed by Zalaletdinov and is detailed in
Refs. \cite{zalaDiffG,mars,zalaMisc}. This scheme, called
Macroscopic Gravity (MG), has been applied to Cosmology in some
specific models \cite{coleyPRL,coley}, where the corrections were
shown to be of the form of a spatial curvature term in the
Friedmann equation. In this paper we address the following
question : Can the equations of Zalaletdinov's MG, in a general
setting and under suitable assumptions, be brought to a form where
the corrections to the FLRW equations can be dealt with in a gauge
independent manner? We will show that the issue of a $3+1$
splitting can be reduced to \emph{one} choice of gauge for the
inhomogeneous metric which  averages to the FLRW metric in its
natural comoving spatial coordinates. While such a choice always
needs to be made for any explicit computations (and is indeed
inherent in the assumption of large scale homogeneity and
isotropy), we will show that it is possible to at least formally
construct \emph{spacetime scalar} quantities which affect the FLRW
equations in a nontrivial manner. We will then  explicitly
construct these scalars assuming that the comoving FLRW spatial
sections result on averaging in a particular gauge choice which
simplifies the formalism, and incidentally also demonstrate that
for this gauge choice, the structure of the corrections in
Zalaletdinov's MG is formally identical to that of the corrections
derived by Buchert \cite{buchDust} (see also
\Cite{coley0704}). Further, we will also show that in principle the
allowed behaviour (with time) of these corrections is, in principle,
more general than of the form of a spatial curvature, as was found in
\Cites{coleyPRL,coley}. 

We have organized the paper as follows : For the benefit of
readers unfamiliar with the details of Macroscopic Gravity, we
have devoted considerable space to recalling this formalism in
Section \ref{recap}, which will also allow us to lay down our
notation. Readers may wish to skim this section and go on to
Section \ref{spatlim}, where we refine the assumptions of standard
Cosmology and adapt them to the framework of MG, and also discuss
the issue of gauge choice. Choosing a particularly convenient
gauge for simplicity, we then explicitly compute the corrections
to the FLRW equations in Section \ref{correln}, and then we
present the formal results for an arbitrary choice of gauge. This
is followed by a comparison between the MG approach and the
averaging scheme developed by Buchert, where we present the case
that in principle the two approaches are expected to be identical
in content, in the sense that they should yield the same physical
results. We conclude in Section \ref{discuss} by summarizing and
discussing some additional issues. \label{intro}

\section{The Macroscopic Gravity Formalism of Zalaletdinov}
\noindent Our aim in this section is to lay out the notation for
the rest of the paper, and to recall the main results and concepts
of Zalaletdinov's Macroscopic Gravity (henceforth MG). The subject
matter we will deal with is of a rather technical nature, and the
results are in fact best presented in the language of differential
forms. We will therefore begin with a brief recollection of some
basic results from differential geometry which will serve to fix
the notation. We will then introduce additional notation while
describing the results of MG. The latter part of this section will
essentially be a repetition of results obtained by Zalaletdinov
\cite{zalaDiffG} and Mars and Zalaletdinov \cite{mars}. While we
shall go through many of the algebraic details in order to make
this paper self-contained, we will not go into details of
arguments justifying the various assumptions made in the averaging
scheme, since our aim here is to apply the formalism developed in
Refs. \cite{zalaDiffG,mars,zalaMisc}.

We shall deal with a differentiable manifold \Cal{M}\ endowed with
a metric $g_{ab}$ of Lorentzian signature $(-+++)$. We will use
lower case Latin characters $a, b, c,\ldots\, i, j, k\ldots\,$ to
denote spacetime indices taking values $0,1,2,3$; later we will
also use upper case Latin characters $A, B, C,...~$ to denote
spatial indices taking values $1,2,3$ in a chosen $3+1$ splitting
of spacetime. We will denote all differential forms and tensors on
\Cal{M} by boldface characters. For example,
$\bm{\alpha}=(1/k!)\alpha_{i_1\cdots
i_k}\dx{i_1}\wedge\cdots\wedge\dx{i_k}$ is a $k$-form written in
the coordinate basis of $1$-forms $\{\dx{i}\}$, while
$\rmb{v}=\rmb{e}_av^a$ is a vector in the basis $\{\rmb{e}_a\}$.
The connection $1$-forms are denoted by $\om^b_{\ph{b}i} =
\left(\Gamma{}^b_{ij}\right)\dx{j}$ where $\Gamma{}^b_{ij}$ are
the Christoffel symbols. For a $k$-form
$\rmb{p}{}^{a\cdots}_{b\cdots}$, we define the ``exterior
covariant derivative'' $\rmb{D}_\omega$ associated with the
connection $\om^a_{\ph{a}b}$, as follows :
\begin{equation}
\rmb{D}_\omega\rmb{p}{}^{a\cdots}_{b\cdots} =
\ext\rmb{p}{}^{a\cdots}_{b\cdots} -
\om^k_{\ph{k}b}\wedge\rmb{p}{}^{a\cdots}_{k\cdots} +
\om^a_{\ph{a}k}\wedge\rmb{p}{}^{k\cdots}_{b\cdots} +\ldots \,.
\label{recap3}
\end{equation}
\noindent The compatibility between the metric and the connection
on \Cal{M}\ is expressed by the condition
\begin{equation}
\rmb{D}_\omega g_{ab} = \ext g_{ab} - g_{ak}\om^k_{\ph{k}b} -
g_{bk}\om^k_{\ph{k}a} = 0 \,, \label{recap4}
\end{equation}
\noindent with a similar condition for the inverse metric
$g^{ab}$. The Cartan structure equations are given by
\begin{subequations}
\begin{equation}
\om^a_{\ph{a}b}\wedge\dx{b}=0 \,, \label{recap5a}
\end{equation}
\begin{equation}
\ext\om^a_{\ph{a}b} + \om^a_{\ph{a}c}\wedge\om^c_{\ph{c}b} =
\rmb{r}^a_{\ph{a}b} \,, \label{recap5b}
\end{equation}
\label{recap5}
\end{subequations}
\noindent where $\rmb{r}^a_{\ph{a}b}$ is the curvature $2$-form on
\Cal{M}\ which defines the Riemann curvature tensor via
$\rmb{r}^a_{\ph{a}b} = (1/2!)r^a_{\ph{a}bcd}\dx{c}\wedge\dx{d}$.
Finally, the structure equations \eqref{recap5} and the metric
compatibility condition \eqref{recap4} are supplemented by their
respective integrability conditions, given by equations
\eqref{recap6a}, \eqref{recap6b} and \eqref{recap6c}
\begin{subequations}
\begin{equation}
\rmb{r}^a_{\ph{a}b}\wedge\dx{b}=0 \,, \label{recap6a}
\end{equation}
\begin{equation}
\ext\rmb{r}^a_{\ph{a}b} - \om^c_{\ph{c}b}\wedge\rmb{r}^a_{\ph{a}c}
+ \om^a_{\ph{a}c}\wedge\rmb{r}^c_{\ph{c}b} =
\rmb{D}_\omega\rmb{r}^a_{\ph{a}b} = 0\,, \label{recap6b}
\end{equation}
\begin{equation}
g_{ak}\rmb{r}^k_{\ph{k}b} + g_{bk}\rmb{r}^k_{\ph{k}a} = 0 \,.
\label{recap6c}
\end{equation}
\label{recap6}
\end{subequations}
We will now proceed to recall the main results of MG. The reader
is referred to Refs. \cite{zalaDiffG,mars} for details of
derivations of the formulae that follow.

The construction of an averaged manifold \Mbar\ from the manifold
\Cal{M}\ begins with the definition of a bilocal exterior calculus
on \Cal{M}, using a bilocal operator \Wxx{a}{j}. Hereafter, the
primed index refers to the point $x^\prime$ and the unprimed index
to the point $x$, and the bivector \Wxx{a}{j}\ transforms like a
vector at $x^\prime$ and a covector at $x$. This bivector is
assumed to be idempotent \cite{mars}, i.e.
$\Cal{W}^{a^\prime}_{c^{\prime\prime}}(x^\prime,x^{\prime\prime})
\Cal{W}^{c^{\prime\prime}}_j(x^{\prime\prime},x) =\Wxx{a}{j}$; and
to have the coincidence limit $\lim_{x^\prime\to x}\Wxx{a}{j} =
\delta{}^a_j$. This ensures that \Wxx{a}{j}\ has the inverse
operator
\Wxxinv{a}{j}. The bivector \Wxx{a}{j}\ serves three purposes -- \\
(a) It is used to ``shift'' the exterior derivative by evaluating
the derivative at $x^\prime$ and antisymmetrizing at $x$ according
to the rule (for say, a $2$-form $\bm{\alpha}$)
\begin{equation}
\ext^\prime_{\Cal{W}}\bm{\alpha}^\prime =
\frac{1}{2!}\alpha_{a^\prime b^\prime,c^\prime}(x^\prime)
\Wxx{c}{j} \dx{j} \wedge \dx{a^\prime} \wedge \dx{b^\prime} \,,
\label{recap7}
\end{equation}
\noindent where $\dx{a^\prime}$ is a coordinate basis $1$-form at
$x^\prime$, and so on. The shifted exterior derivative is used to
define a fully bilocal exterior derivative $\dbil$ as
\begin{equation}
\dbil = \ext + \ext^\prime_{\Cal{W}}\,. \label{recap8}
\end{equation}
\noindent (b) \W{a}{j} is used to define the bilocal extension of
tensors and $k$-forms. For example, if $P{}^a_b(x)$ is a $(1,1)$
tensor on \Cal{M}, then its bilocal extension is defined as
\begin{equation}
\bil{P}{}^a_b(x^\prime,x) =
\Wxx{b}{b}\Wxxinv{a}{a}P{}^{a^\prime}_{b^\prime}(x^\prime) \,.
\label{recap9}
\end{equation}
\noindent Similarly, if $\bm{\alpha}$ is a $2$-form on \Cal{M},
its bilocal extension is defined as
\begin{equation}
\bil{\bm{\alpha}}(x^\prime,x) = \frac{1}{2!}\alpha_{a^\prime
  b^\prime}(x^\prime)\Wxx{a}{j}\Wxx{b}{k} \dx{j}\wedge\dx{k}\,.
\label{recap10}
\end{equation}
\noindent The definitions above can be easily generalised for an
arbitrary tensor-valued $k$-form on \Cal{M}. Using the definition
of bilocal extensions, the averages of say $P{}^a_b$ and
$\bm{\alpha}$ over a spacetime region $\bm{\Sigma}$ with a
supporting point $x$, are then given by
\begin{align}
&\bar P{}^a_b(x) = \stavg{\bil{P}{}^a_b} = \frac{1}{V_\Sigma}
\int_\Sigma{d^4x^\prime\sqrt{-g^\prime}\bil{P}{}^a_b(x^\prime,x)}
~;\nonumber\\
&\bar{\bm{\alpha}}(x) =
\frac{1}{V_\Sigma}\int_\Sigma{d^4x^\prime\sqrt{-g^\prime}
  \bil{\bm{\alpha}}(x^\prime,x)} ~~;~~ V_\Sigma =
\int_\Sigma{d^4x^\prime\sqrt{-g^\prime}} \,, \label{recap11}
\end{align}
the subscript $ST$ standing for `spacetime'. (Note that
$\bar{\bm{\alpha}}(x)$ is a (local) $k$-form at $x$.) It can be
shown \cite{zalaDiffG,mars} that a necessary and sufficient
condition for the bilocal exterior derivative to be nilpotent
($\dbil\dbil=0$) and hence for the average of a tensor-valued
$k$-form to be a single valued function of the supporting point
$x$, is given by
\begin{equation}
\dbil\rmb{W}^{a^\prime} = 0 ~~;~~ \rmb{W}^{a^\prime} =
\W{a}{i}\dx{i}\,, \label{recap12}
\end{equation}
\noindent and further \cite{mars} that a bivector \W{a}{j}\
satisfying all the required properties always exists on an
$n$-dimensional differentiable
manifold with a volume $n$-form. \\
(c) Finally, to completely define the exterior derivatives of
averaged quantities, \W{a}{j}\ is used to specify a Lie dragging
of the averaging region $\bm{\Sigma}$. This ensures that the
volumes of the averaging regions constructed at nearby supporting
points are coordinated in a well defined manner. Suppose $x^a$ and
$x^a+\xi^a\Delta\lambda$ are the coordinates of two support
points, where $\Delta\lambda$ is a small change in the parameter
along the integral curve of a given vector field $\xi^a$.
Symbolically denoting the two points as $x$ and
$x+\xi\Delta\lambda$, the averaging region at $x+\xi\Delta\lambda$
is defined in terms of the averaging region $\bm{\Sigma}$ at $x$,
by transporting every point $x^\prime\in\bm{\Sigma}$ around $x$
along the appropriate integral curve of a \emph{new} bilocal
vector field $S^{a^\prime}$ defined as $S^{a^\prime}(x^\prime,x) =
\Wxx{a}{j}\xi^j(x)$, and thereby constructing the averaging region
$\bm{\Sigma}(\Delta\lambda)$ with support point
$x+\xi\Delta\lambda$. (See Refs. \cite{zalaDiffG,mars} for further
discussion on the significance of this averaging region
coordination.) With this definition, the exterior derivative of
the average of a $k$-form $\rmb{p}{}^a_b$ can be written as
\cite{zalaDiffG}
\begin{equation}
\ext \bar{\rmb{p}}{}^a_b = \stavg{\dbil\bil{\rmb{p}}{}^a_b} +
\stavg{{\rm div}_\epsilon \rmb{W}\wedge\bil{\rmb{p}}{}^a_b} -
\stavg{{\rm div}_\epsilon \rmb{W}}\wedge\bar{\rmb{p}}{}^a_b\,,
\label{recap13}
\end{equation}
where we have defined, in keeping with the notation of
\Cite{zalaDiffG},
\begin{align}
{\rm div}_\epsilon \rmb{W} &=
 \Cal{W}{}^{a^\prime}_{j:a^\prime}\dx{j}\nonumber\\
 &\equiv \left(\Cal{W}{}^{a^\prime}_{j,a^\prime} +
 \W{a}{j}\partial_{a^\prime}\ln\sqrt{-g^\prime} \right)\dx{j} \,.
\label{recap14}
\end{align}
\noindent Clearly, it is desirable to choose a coordination
bivector \W{a}{j} which satisfies
\begin{equation}
{\rm div}_\epsilon \rmb{W}=0\,, \label{recap15}
\end{equation}
\noindent since firstly, this allows us to commute the exterior
derivative with the averaging according to
\begin{equation}
\ext \bar{\rmb{p}}{}^a_b = \stavg{\dbil\bil{\rmb{p}}{}^a_b}\,,
\label{recap16}
\end{equation}
\noindent and secondly, it implies that the volume of the
averaging region is held constant during the coordination
\cite{mars}, and is therefore a free parameter in the formalism.
It can be shown \cite{mars} that firstly, the general solution of
\eqn{recap12} for an idempotent coordination bivector is given by
\begin{equation}
\Wxx{a}{j} = f^{a^\prime}_m(x^\prime) f^{-1\,m}_{\ph{-1}j}(x) \,,
\label{recap17}
\end{equation}
\noindent where $f^a_m(x)\bm{\partial}_a = \rmb{f}_m$ is any
vector basis satisfying the commutation relations
\begin{equation}
\left[\rmb{f}_i,\rmb{f}_j\right] = C{}^k_{ij}\rmb{f}_k ~~;~~
C{}^k_{ij} = {\rm constant} \,, \label{recap18}
\end{equation}
\noindent and secondly, that \eqn{recap15} with the coordination
bivector given by \eqn{recap17} is always integrable on a
differentiable manifold with a given volume $n$-form. Further, for
the special class of bivectors for which $C{}^k_{ij} \equiv 0$,
the vectors $\{\rmb{f}_k\}$ form a coordinate basis, with `proper'
coordinate functions $\phi^m(x)$ say, so that
\begin{equation}
f^{a}_m(x(\phi^n)) = \frac{\partial x^a}{\partial\phi^m} ~~;~~
f^{-1\,m}_{\ph{-1}j}(\phi(x^k)) = \frac{\partial \phi^m}{\partial
x^j} \,, \label{recap19}
\end{equation}
\noindent and satisfying \eqn{recap15} makes this proper
coordinate system volume preserving, with $g(\phi^m)=\,$constant.
When expressed in terms of such a volume preserving coordinate
(VPC) system, the coordination bivector takes its most simple
form, namely
\begin{equation}
\Wxx{a}{j}\mid_{\rm proper} = \delta{}^{a^\prime}_j\,.
\label{recap20}
\end{equation}
\noindent 
Volume preserving coordinates in fact form a large class in
themselves, generalizing the Cartesian coordinate system of Minkowski
spacetime. For a discussion on the properties of VPCs and the
associated bivectors \W{a}{a}, see Sec. 8 of \Cite{mars}. We now turn
to describing the geometry on a manifold \Mbar\ which is to be
considered an averaged version of the manifold \Cal{M}. We denote the
bilocal extension of the connection $1$-form $\om^a_{\ph{a}b}$ on
\Cal{M}, as $\Om{a}{b}$, so that 
\begin{equation}
\Om{a}{b}(x^\prime,x) = \Gamma{}^{a^\prime}_{b^\prime
  c^\prime}(x^\prime) \Wxxinv{a}{a}\Wxx{b}{b}\Wxx{c}{c}\dx{c}\,.
\label{recap21}
\end{equation}
\noindent The key idea of MG is that the average $\barOm{a}{b}$ of
the connection $1$-form on \Cal{M}, is to be considered the
connection $1$-form on the averaged manifold \Mbar. The goal is
then to average out the bilocal extensions of the structure
equations \eqref{recap5} and the compatibility condition
\eqref{recap4} and their integrability conditions \eqref{recap6},
and to express them in terms of appropriate differential forms
defined on \Mbar. The bilocal extensions of Eqns. \eqref{recap5}
and \eqref{recap4}, are respectively given by
\begin{subequations}
\begin{equation}
\Om{a}{b}\wedge\dx{b}=0 \,, \label{recap22a}
\end{equation}
\begin{equation}
\dbil\Om{a}{b} + \Om{a}{c}\wedge\Om{c}{b} =
\bil{\rmb{r}}^a_{\ph{a}b} \,, \label{recap22b}
\end{equation}
\begin{equation}
\Dbil{\Omega} \bil{g}_{ab} = \dbil \bil{g}_{ab} -
\bil{g}_{ak}\Om{k}{b} - \bil{g}_{bk}\Om{k}{a} = 0 \,,
\label{recap22c}
\end{equation}
\label{recap22}
\end{subequations}
\noindent where, in the last equation, $\Dbil{\Omega}$ is the
bilocal covariant exterior derivative associated with the bilocal
connection $\Om{a}{b}$. The integrability conditions of
\eqns{recap22} are given by the bilocal extensions of
\eqns{recap6},
\begin{subequations}
\begin{equation}
\bil{\rmb{r}}^a_{\ph{a}b}\wedge\dx{b}=0 \,, \label{recap23a}
\end{equation}
\begin{equation}
\dbil\bil{\rmb{r}}^a_{\ph{a}b} -
\Om{c}{b}\wedge\bil{\rmb{r}}^a_{\ph{a}c} +
\Om{a}{c}\wedge\bil{\rmb{r}}^c_{\ph{c}b} = 0 \,, \label{recap23b}
\end{equation}
\begin{equation}
\bil{g}_{ak}\bil{\rmb{r}}^k_{\ph{k}b} +
\bil{g}_{bk}\bil{\rmb{r}}^k_{\ph{k}a} = 0 \,. \label{recap23c}
\end{equation}
\label{recap23}
\end{subequations}
\noindent To proceed with the averaging, a correlation $2$-form is
defined
\begin{equation}
\bZ{a}{b}{i}{j} = \stavg{\Om{a}{b}\wedge\Om{i}{j}} -
\barOm{a}{b}\wedge\barOm{i}{j}\,. \label{recap24}
\end{equation}
\noindent The average of the curvature $2$-form
$\rmb{r}^a_{\ph{a}b}$ on \Cal{M}\ is denoted
$\rmb{R}^a_{\ph{a}b}\equiv\stavg{\,\bil{\rmb{r}}^a_{\ph{a}b}}$ ,
and the curvature $2$-form on the averaged manifold \Mbar\ is
denoted $\rmb{M}^a_{\ph{a}b}$,
\begin{equation}
\rmb{M}^a_{\ph{a}b} = \ext\barOm{a}{b} +
\barOm{a}{k}\wedge\barOm{k}{b}\,. \label{recap25}
\end{equation}
\noindent Equations \eqref{recap22a} and \eqref{recap22b} then
average out to give
\begin{subequations}
\begin{equation}
\barOm{a}{c}\wedge\dx{c} = 0\,, \label{recap26a}
\end{equation}
\begin{equation}
\rmb{M}^a_{\ph{a}b} = \rmb{R}^a_{\ph{a}b} - \bZ{a}{k}{k}{b}\,.
\label{recap26b}
\end{equation}
\label{recap26}
\end{subequations}
\noindent It can further be shown \cite{zalaDiffG} that the
algebraic identities of the curvature $2$-form on \Mbar\ hold :
\begin{equation}
\rmb{M}^a_{\ph{a}c}\wedge\dx{c} = 0 ~~;~~ \rmb{M}^a_{\ph{a}a} =
0\,, \label{recap27}
\end{equation}
\noindent To average out \eqn{recap23b}, one needs to introduce a
correlation $3$-form which fixes the differential properties of
the $2$-form $\bZ{a}{b}{i}{j}$. The differential properties of
this $3$-form are then fixed by introducing a $4$-form. However,
accounting for these additional correlation forms is extremely
complicated. We will therefore make the simpler assumption
(consistent with the formalism) of setting the correlation
$3$-form and $4$-form to zero, with the condition
\begin{equation}
\rmb{D}_{\bar\Omega} \bZ{a}{b}{i}{j} = 0\,.
\label{recap28} 
\end{equation}
\noindent 
It will be an interesting exercise to check whether this assumption
is, in fact, justified in case of averaging an exact inhomogeneous
model. \eqn{recap28} has the integrability condition \cite{zalaDiffG} 
\begin{equation}
\mathbb{P}\left(\rmb{R}^a_{\ph{a}c}\wedge\bZ{c}{b}{i}{j} -
\bZ{a}{b}{i}{k}\wedge\rmb{R}^k_{\ph{k}j} \right) = 0\,.
\label{recap29}
\end{equation}
\noindent 
Here the symbol $\mathbb{P}$ permutes the free indices in, say
$\rmb{K}^{a\ph{b}i\ph{j}m}_{\ph{a}b\ph{i}j\ph{m}n}$ pairwise according
to $\mathbb{P}(\rmb{K}^{a\ph{b}i\ph{j}m}_{\ph{a}b\ph{i}j\ph{m}n}) = 
(1/3!)(\rmb{K}^{a\ph{b}i\ph{j}m}_{\ph{a}b\ph{i}j\ph{m}n} -
\rmb{K}^{i\ph{j}a\ph{b}m}_{\ph{i}j\ph{a}b\ph{m}n} +
\rmb{K}^{i\ph{j}m\ph{n}a}_{\ph{i}j\ph{m}n\ph{a}b} - \ldots)$, and
any summed indices are ignored. Setting the correlation $3$-form
and $4$-form to zero also imposes the condition
\cite{zalaMisc,coleyPRL}
\begin{equation}
\mathbb{P}\left(\bZ{a}{b}{c}{d} \wedge \bZ{d}{i}{j}{k} \right) = 0
\,. \label{recap29add}
\end{equation}
\noindent \eqn{recap23b} averages out to give the Bianchi
identities for the curvature $2$-form on \Mbar
\begin{equation}
\rmb{D}_{\bar\Omega}\rmb{M}^a_{\ph{a}b} = \ext\rmb{M}^a_{\ph{a}b}
- \barOm{k}{b}\wedge\rmb{M}^a_{\ph{a}k} +
\barOm{a}{k}\wedge\rmb{M}^k_{\ph{k}b} = 0\,. \label{recap30}
\end{equation}
\noindent To average out equations \eqref{recap22c} and
\eqref{recap23c}, one needs to make additional assumptions. For a
class of slowly varying tensor fields (tensor-valued $k$-forms)
$\rmb{c}{}^{m\cdots}_{n\cdots}$ on \Cal{M}\ such as the metric and
other covariantly constant tensors, and Killing tensors, etc., if
one assumes that
\begin{subequations}
\begin{equation}
\stavg{\Om{a}{b}\wedge\bil{\rmb{c}}{\,}^{m\cdots}_{n\cdots}} =
\barOm{a}{b}\wedge\bar{\rmb{c}}{\,}^{m\cdots}_{n\cdots} \,,
\label{recap31a}
\end{equation}
\begin{equation}
\stavg{\Om{a}{b}\wedge\Om{i}{j}\wedge
  \bil{\rmb{c}}{\,}^{m\cdots}_{n\cdots}} =
  \stavg{\Om{a}{b}\wedge\Om{i}{j}}\wedge
  \bar{\rmb{c}}{\,}^{m\cdots}_{n\cdots} \,,
\label{recap31b}
\end{equation}
\label{recap31}
\end{subequations}
\noindent then \eqn{recap22c} (and its analogue for $g^{ab})$
average out to give
\begin{equation}
\rmb{D}_{\bar\Omega} \bar g_{ab} = 0 ~~;~~ \rmb{D}_{\bar\Omega}
\bar g^{ab} = 0 \,. \label{recap32}
\end{equation}
\noindent Further, for a general slowly varying object
$\rmb{c}{}^{m\cdots}_{n\cdots}$, the following identity holds
\begin{align}
&\stavg{\bil{\rmb{r}}{}^a_{\ph{a}b} \wedge
  \bil{\rmb{c}}{}^{m\cdots}_{n\cdots}} - \rmb{R}^a_{\ph{a}b} \wedge
  \bar{\rmb{c}}{}^{m\cdots}_{n\cdots} \nonumber\\
 &- \stavg{\Om{a}{b} \wedge
  \Dbil{\Omega}\bil{\rmb{c}}{}^{m\cdots}_{n\cdots}} + \barOm{a}{b}
  \wedge \rmb{D}_{\bar\Omega} \bar {\rmb{c}}{}^{m\cdots}_{n\cdots}
  \nonumber\\ 
 &= -\bZ{a}{b}{m}{j} \wedge \bar{\rmb{c}}{}^{j\cdots}_{n\cdots}
  -\ldots +  \bZ{a}{b}{j}{n} \wedge  \rmb{c}{}^{m\cdots}_{j\cdots}
  +\ldots~~\,, 
\label{recap33}
\end{align}
\noindent which averages out \eqn{recap23c} (and its analogue for
$g^{ab}$) to give
\begin{equation}
\bar g_{ak}\rmb{M}^k_{\ph{k}b} + \bar g_{kb}\rmb{M}^k_{\ph{k}a} =
0 ~~;~~ \rmb{M}^a_{\ph{a}k}\bar g^{kb} + \rmb{M}^a_{\ph{a}k}\bar
g^{kb} = 0\,. \label{recap34}
\end{equation}
\noindent \eqn{recap32} allows one to choose $G_{ab} = \bar
g_{ab}$, where $G_{ab}$ is the metric on the averaged manifold
\Mbar. In general however, we have $G^{ab}\neq\bar g^{ab}$, and
one defines the tensor $U^{ab}\equiv \bar g^{ab}-G^{ab}$ to keep
track of this difference. (See \Cite{zalaDiffG} for details.)
However, we show in Appendix \eqref{app-gbar} that when the
averaged manifold is highly symmetric, as in the case of a
manifold with homogeneous and isotropic spatial sections which we
will consider, one finds that $U^{ab}=0$ (see also the last paper
in \Cite{zalaMisc}). In the general case, it turns out that
\eqn{recap33} is all that is needed to average out the Einstein
equations
\begin{equation}
g^{ak}r_{kb} - \frac{1}{2}\delta{}^a_b g^{ij}r_{ij} = -\kappa
t{}^{a({\rm mic})}_b\,, \label{recap35}
\end{equation}
\noindent where $\kappa = 8\pi G_N$, $t{}^{a({\rm mic})}_b$ is the
microscopic energy momentum tensor of the matter distribution, and
the Ricci tensor $r_{ab}$ on \Cal{M}\ is defined according to the
sign convention $r_{ab} = r^j_{\ph{j}abj}$. The averaging leads to
the equations
\begin{align}
G^{ak}M_{kb} - \frac{1}{2}\delta{}^a_b G^{ij}M_{ij} &= -\kappa
\stavg{\bil{t}{\,}^{a({\rm mic})}_b} \nonumber\\ 
&+ \left(Z^a_{\ph{a}ijb} -
\frac{1}{2}\delta{}^a_b Z^k_{\ph{k}ijk}\right)\bar g^{ij} \nonumber\\ 
&- \left(U^{ak}M_{kb} -
\frac{1}{2}\delta{}^a_bU^{ij}M_{ij}\right)\,, 
\label{recap36}
\end{align}
\noindent where $M_{ab} = M^j_{\ph{j}abj}$ is the Ricci tensor on
\Mbar\ and we have defined
\begin{equation}
Z^a_{\ph{a}ijb} = 2 Z^{a\ph{ik}k}_{\ph{a}ik\ph{k}jb} ~~;~~
  \bZ{a}{b}{i}{j} = Z^{a\ph{bm}i}_{\ph{a}bm\ph{i}jn}\dx{m}\wedge\dx{n}
  \,.
\label{recap37}
\end{equation}
\noindent The averaged equations \eqref{recap36} differ from the
usual Einstein equations by the correlation tensor which we define
as
\begin{equation}
C{}^a_b = \left(Z^a_{\ph{a}ijb} - \frac{1}{2}\delta{}^a_b
Z^m_{\ph{a}ijm} \right)\bar g^{ij} - \left(U^{ak}M_{kb} -
\frac{1}{2}\delta{}^a_bU^{ij}M_{ij}\right) \,. \label{recap38}
\end{equation}
\noindent Hence, denoting the Einstein tensor on \Mbar\ as
$E{}^a_b$, and defining the tensor $T{}^a_b$ via
\begin{equation}
T{}^a_b = \stavg{\bil{t}{\,}^{a({\rm mic})}_b}\,, \label{recap39}
\end{equation}
\noindent the averaged Einstein equations read
\begin{equation}
E{}^a_b = -\kappa T{}^a_b + C{}^a_b\,. \label{recap40}
\end{equation}
\noindent Since the left hand side of \eqn{recap40} is covariantly
conserved by construction ($E{}^a_{b;a} = 0$), where the semicolon
denotes covariant differentiation with respect to the connection
on \Mbar, in general one has
\begin{equation}
\left(-\kappa T{}^a_b + C{}^a_b\right)_{;a} = 0\,, \label{recap41}
\end{equation}
\noindent with no condition on $T{}^a_b$ and $C{}^a_b$ separately.
However, the condition \eqref{recap28} after taking appropriate
traces, reduces to
\begin{equation}
C{}^a_{b;a} = 0\,, \label{recap42}
\end{equation}
\noindent (see the second of \Cite{zalaMisc}) which implies that
the averaged energy-momentum tensor $T{}^a_b$ is also covariantly
conserved.

It can also be shown that in 4 dimensions, the 720 \emph{a priori}
independent components of $Z^{a\ph{bm}i}_{\ph{a}bm\ph{i}jn}$ are
subject to 680 constraints arising from \eqns{recap29} and
\eqref{recap29add}. This leaves 40 independent components which
combine to give the 10 independent components of the correlation
tensor $C{}^a_b$. The conditions in \eqns{recap29} and
\eqref{recap29add} do not constrain the components of $C{}^a_b$,
which follows from considering the structure of those equations.
(See also \Cite{coleyPRL}.)
\label{recap}

\section{A $3+1$ spacetime splitting and the Spatial Averaging limit}
\noindent We are now in a position to apply the MG formalism to
the problem of Cosmology. We start with the assumption that
Einstein's equations are to be imposed on length scales where
stars are pointlike objects (we denote such a scale as \Linh). The
averaging we perform will be directly at a length scale \Lfrw\
larger than about $100h^{-1}$Mpc or so. This averaging scale is
assumed to satisfy $\Linh\ll\Lfrw\ll\Lhub$ where \Lhub\ is the
length scale of the observable universe. The averaging will be
assumed to yield a geometry which has homogeneous and isotropic
spatial sections. In other words, we will assume that the averaged
manifold \Mbar\ admits a preferred, hypersurface-orthogonal unit
timelike vector field $\bar v^a$, which defines $3$-dimensional
spacelike hypersurfaces of constant curvature, and that $\bar v^a$
is tangent to the trajectories of observers who see an isotropic
Cosmic Background Radiation. (These ``observers'' are defined in
the averaged manifold -- we will clarify below what they
correspond to in the inhomogeneous manifold.) Throughout the rest
of this paper, for simplicity, we will work with the special case
where the spatial sections on \Mbar\ defined by $\bar v^a$ are
flat. (In principle the entire calculation can be repeated for
non-flat spatial sections as well.) One can then choose
coordinates $(t,x^A)$, $A=1,2,3$, on \Mbar\ such that the spatial
line element takes the form
\begin{equation}
^{(\Mbar)}ds^2_{\rm spatial} = a^2(t)\delta_{AB}dx^Adx^B\,,
\label{spatlim1}
\end{equation}
where $\delta_{AB}=1$ for $A=B$, and $0$ otherwise, and we have
$\bar v^a=(\bar v^t,0,0,0)$ so that the spatial coordinates are
comoving with the preferred observers. The vector field $\bar v^a$
also defines a proper time (the cosmic time) $\tau$ such that
$\partial_\tau = \bar v^a\partial_a = \bar v^t\partial_t$. We will
further assume that the averaged energy-momentum tensor $T{}^a_b$
can be written in the form of a perfect fluid, as
\begin{equation}
T{}^a_b = \rho \bar v^a \bar v_b + p\pi^a_b\,, \label{spatlim2}
\end{equation}
where the projection operator $\pi^a_b$ is defined as
\begin{equation}
\pi^a_b = \delta{}^a_b + \bar v^a\bar v_b \,, \label{spatlim3}
\end{equation}
and $\rho$ and $p$ are the homogeneous energy density and pressure
respectively, as measured by observers moving on trajectories (in
\Mbar) with the tangent vector field $\bar v^a$,
\begin{equation}
\rho\equiv T{}^a_b\bar v^b\bar v_a ~~;~~ p\equiv\frac{1}{3}\pi^b_a
T{}^a_b\,. \label{spatlim-T-ab1}
\end{equation}
$\rho$ and $p$ are observationally relevant quantities, since all
measurements of the matter energy density, especially those from
studies of Large Scale Structure, interpret observations in the
context of the averaged geometry. (See Refs.
\cite{buch-carf,wiltshire} for more careful treatments of this
point.) An important consequence of the above assumptions is that
the correlation tensor $C{}^a_b$, when expressed in terms of the
natural coordinates adapted to the spatial sections defined by the
vector field $\bar v^a$, is \emph{spatially homogeneous}. This is
clear when the modified Einstein equations \eqref{recap40} are
written in these natural coordinates.

The existence of the vector field $\bar v^a$ with the attendant
assumptions described above, allows us to separate out the
nontrivial components of the (FLRW) Einstein tensor $E{}^a_b$ on
\Mbar\ in a coordinate independent fashion -- the Einstein tensor
can be written as
\begin{align}
E{}^a_b = j_1(x) \bar v^a\bar v_b &+ j_2(x) \pi^a_b \nonumber\\
j_1(x) \equiv E{}^a_b\bar v^b\bar v_a ~~&;~~ j_2(x) \equiv
\frac{1}{3}\left( \pi^b_a E{}^a_b\right)\,, \label{spatlim-FLRW1}
\end{align}
where $j_1(x)$ and $j_2(x)$ are scalar functions whose form
depends upon the coordinates used. The remaining components given
by $\pi^b_k E{}^a_b\bar v_a$ and the traceless part of
$\pi^i_a\pi^b_k E{}^a_b$, vanish identically. Since the
energy-momentum tensor $T{}^a_b$ in \eqn{spatlim2} also has an
identical structure, this structure is therefore \emph{also
imposed} on the correlation tensor $C{}^a_b$. Namely, $\pi^b_k
C{}^a_b\bar v_a$ and the traceless part of $\pi^i_a\pi^b_k
C{}^a_b$ \emph{must vanish}. This is a condition on the underlying
inhomogeneous geometry, irrespective of the coordinates used on
either \Cal{M}\ or \Mbar, and is clearly a consequence of
demanding that the averaged geometry have the symmetries of the
FLRW spacetime. (Alternatively, one could model the averaged
energy-momentum tensor as having some anisotropic components,
which would then have to be balanced by the corresponding
components of the correlation tensor. We will not pursue this idea
here, and will restrict ourselves to the more conservative
assumption of a homogeneous and isotropic averaged matter
distribution \eqref{spatlim2}.)

This leads us to the crucial question of the choice of
\emph{gauge} for the underlying geometry : namely, what choice of
spatial sections for the \emph{inhomogeneous} geometry, will lead
to the spatial sections of the FLRW metric in the comoving
coordinates defined in \eqn{spatlim1}? Since the matter
distribution at scale \Linh\ need not be pressure-free (or,
indeed, even of the perfect fluid form), there is clearly no
natural choice of gauge available, although locally, a synchronous
reference frame can always be chosen. We note that there must be
\emph{at least one} choice of gauge in which the averaged metric
has spatial sections in the form \eqref{spatlim1} -- this is
simply a refinement of the Cosmological Principle, and of the Weyl
postulate, according to which the Universe is homogeneous and
isotropic on large scales, and individual galaxies are considered
as the ``observers'' travelling on trajectories with tangent $\bar
v^a$. In the averaging approach, it makes more sense to replace
``individual galaxies'' with the \emph{averaging domains}
considered as physically infinitesimal cells -- the ``points'' of
the averaged manifold \Mbar. This is physically reasonable since
we know after all, that individual galaxies exhibit peculiar
motions, undergo mergers and so on. This idea is also more in
keeping with the notion that the Universe is homogeneous and
isotropic \emph{only on the largest
  scales}, which are much larger than the scale of individual
galaxies. (See also \Cite{mars} for a discussion of how this
assumption of treating the averaging domains as being effectively
point-like, is essential for the idempotency of the averaging
operation.)

Consider any $3+1$ spacetime splitting in the form of a lapse
function ${N}(t,x^J)$, a shift vector ${N}^A(t,x^J)$, and a metric
for the $3$-geometry ${h}_{AB}(t,x^J)$, so that the line element
on \Cal{M}\ can be written as
\begin{equation}
^{(\Cal{M})}ds^2 = -\left({N}^2 -
  {N}_{\!A}{N}^A\right)dt^2 + 2{N}_{\!B}dx^Bdt +
  {h}_{AB}dx^Adx^B\,,
\label{spatlim4}
\end{equation}
where ${N}_{\!A} = {h}_{\!AB}{N}^B$. At first sight, it might seem
reasonable to leave the choice of gauge arbitrary. One could then
formally consider a coordination bivector given by the
\eqns{recap17} and \eqref{recap19}, with $x^i$ denoting the
coordinates in the chosen gauge and $\phi^m$ the VPCs; and demand
for example, that the metric \eqref{spatlim4} (with say ${N}^A=0$)
average out to the FLRW form (with a nonsynchronous time
coordinate in general). This would imply
\begin{align}
G_{00} = \stavg{\bil{g}_{00}} &= -f^2(t) ~~;~~ G_{0A} =
\stavg{\bil{g}_{0A}} = 0 ~;\nonumber\\
 G_{AB} &= \stavg{\bil{g}_{AB}} =
a^2(t)\delta_{AB} \,. \label{spatlim-gauge1}
\end{align}
There is no \emph{a priori} reason to assume that the functions
$f(t)$ and $a(t)$ are related. Note that the condition on the
bilocal extension $\bil{g}_{0A}(x^\prime,x)$ is in general
nontrivial even when the components $g_{0A}(x)$ are chosen to be
zero. In the Appendix \eqref{app-gbar} we show that with the above
assumptions, for a general lapse function ${N}$, the conditions
$\rmb{D}_{\bar\Omega}\bar g^{ab}=0$ (\eqn{recap32}) also allow us
to choose
\begin{equation}
U^{ij}\equiv\bar g^{ij}-G^{ij}=0\,. \label{spatlim-gauge2}
\end{equation}
However, it turns out that for a general lapse function ${N}$, the
explicit form of the correlation terms in the averaged equations
with the assumptions \eqref{spatlim-gauge1}, is rather complicated
(although one can write down these terms formally in a relatively
compact manner). On the other hand, if we make the assumption that
the spatial sections on \Cal{M}\ leading to the spatial metric
\eqref{spatlim1} on \Mbar, are spatial sections \emph{in a volume
  preserving gauge}, then the correlation terms simplify
greatly. This is not surprising since the MG formalism is nicely
adapted to the choice of volume preserving coordinates.

We therefore adopt the following procedure : We will first make
the set of assumptions which allow us to use a volume preserving
gauge. Using these assumptions in the remainder of this section,
we will introduce the notion of spatial averaging within the MG
framework. In Section \ref{correln} we will calculate the
correlation terms and display the modified equations resulting
from this particular choice of gauge. This exercise can be
thought of as a toy calculation with simplifying assumptions.
Following that calculation we will show how the correlation terms
can be generalized to the more physically relevant case where the
gauge in the inhomogeneous metric is (formally) left unspecified.

To begin our first calculation, we perform a coordinate
transformation and shift to the gauge wherein the new lapse
function $N$ is given by $N=1/\sqrt{h}$ where $h$ is the
determinant of the new $3$-metric $h_{AB}$. In general, one will
now be left with a non-zero shift vector $N^A$; however, the
condition $N\sqrt{h}=1$ ensures that the coordinates we are now
using are volume preserving, since the metric determinant is given
by $g=-N^2h=-1={\rm constant}$. We denote these volume preserving
coordinates (VPCs) by $(\bt, \rmb{x}) = (\bt, x^A) = (\bt, x, y,
z)$, and will assume that the spatial coordinates are non-compact.
For simplicity, we make the added assumption that $N^A=0$ in the
inhomogenous geometry \cite{note1}, so that
$g_{\bt\,\bt}=-N^2=-1/h$ and $g_{\bt A}=0$. The line element for
the inhomogenous manifold \Cal{M}\ becomes
\begin{equation}
^{(\Cal{M})}ds^2=-\frac{d\bt^2}{h(\bt,\rmb{x})} +
  h_{AB}(\bt,\rmb{x})dx^Adx^B\,.
\label{spatlim5}
\end{equation}
Note that in this gauge, the average takes on a particularly
simple form : for a tensor $p{}^i_j(x)$, with a spacetime
averaging domain given by the ``cuboid'' $\bm{\Sigma}$ defined by
\begin{equation}
\bm{\Sigma} =
\left\{(\bt,x,y,z)\mid-T/2<\bt<T/2,-L/2<x,y,z<L/2\right\},
\label{spatlim6}
\end{equation}
where $T$ and $L$ are averaging time and length scales
respectively, the average is given by
\begin{align}
&\stavg{\bil{p}{\,}^i_j}(\bt,\rmb{x}) =
    \stavg{p{}^i_j}(\bt,\rmb{x}) \nonumber\\ 
&=  \frac{1}{TL^3}
    \int_{\bt-T/2}^{\bt+T/2}{dt^\prime\int_{-L/2}^{+L/2}{
    dx^\prime dy^\prime dz^\prime\bigg[
      p{}^i_j(t^\prime,x^\prime,y^\prime,z^\prime)\bigg]}} \,,
\label{spatlim7}
\end{align}
where the limits on the spatial integral are understood to hold for
all three spatial coordinates. We define the ``spatial averaging
limit'' as the limit $T\to0$ (or $T\ll\Lhub$) which is interpreted as
providing a definition of the average on a spatial domain
corresponding to a ``thin'' time slice, the averaging operation now
being given by 
\begin{align}
&\avg{p{}^i_j}(\bt,\rmb{x}) \nonumber\\
&=  \frac{1}{L^3}
    \int_{-L/2}^{+L/2}{dx^\prime dy^\prime dz^\prime\bigg[ 
      p{}^i_j(\bt,x^\prime,y^\prime,z^\prime)\bigg]} +
    \Cal{O}\left(TL_{\rm Hubble}^{-1} \right) \,.
\label{spatlim8}
\end{align}
(Note the time dependence of the integrand.) Henceforth, averaging
will refer to spatial averaging, and will be denoted by
$\avg{...}$, in contrast to the spacetime averaging considered
thus far (denoted by $\stavg{...}$). The choice of a cube with
sides of length $L$ as the spatial averaging domain was arbitrary,
and is in fact not essential for any of the calculations to
follow. In particular, all calculations can be performed with a
spatial domain of arbitrary shape \cite{note2}. We will only use
the cube for definiteness and simplicity in displaying equations.
(An assumption essentially amounting to the limit in
\eqn{spatlim8} was also made in \Cite{coley}, in the context of
averaging the spherically symmetric and inhomogeneous
Lema\^itre-Tolman-Bondi models.) The significance of introducing a
spatial averaging in this manner is that the construction of
spatial averaging is not isolated from spacetime averaging, but is
a special limiting case of the latter and is, in fact, still a
fully covariant operation.


For the volume preserving gauge, we now make the averaging
assumption \eqref{spatlim-gauge1} (with the averaging scale
$L=\Lfrw$), which reduces to
\begin{align}
G_{\bt\bt} &= \avg{g_{\bt\,\bt}} = \avg{\frac{-1}{h}} = -f^2(\bt) ~;
\nonumber\\ 
&G_{AB} = \avg{h_{AB}}=\ba^2(\bt)\delta_{AB} \,,
\label{spatlim9}
\end{align}
where $\ba$ and $f$ are some functions of the time coordinate
alone. A few remarks are in order on this particular choice of
assumptions. Apart from the fact that the spacetime averaging
operation takes on its simplest possible form \eqref{spatlim7} in
this gauge and allows a transparent definition of the spatial
averaging limit, it can also be shown that the assumptions in
\eqn{spatlim9} are sufficient to establish the following relations
:
\begin{equation}
f^2(\bt) = \avg{\frac{1}{h}} = \frac{1}{\avg{h}} =
\frac{1}{\ba^6}\,. \label{spatlim10}
\end{equation}
Here the second equality arises from the condition $\bar
g^{ij}=G^{ij}$ which can be assumed whenever the averaged metric
is of the FLRW form (see Appendix \eqref{app-gbar}). The last
equality follows on considering the conditions
$\avg{\bil{\Gamma}{}^a_{bc}} =\,^{(\rm FLRW)}\Gamma{}^a_{bc}$ in
obvious notation, (the basic assumption of the MG averaging
scheme), details of which can be found in the Appendix
\eqref{app-gamma}. \eqn{spatlim10} reduces the line element on
\Mbar\ to the form
\begin{equation}
^{(\Mbar)}ds^2 = -\frac{d\bt^2}{\ba^6(\bt)} +
  \ba^2(\bt)\delta_{AB}dx^Adx^B \ .
\label{spatlim11}
\end{equation}
The line element in \eqn{spatlim11} clearly corresponds to the
FLRW metric in a \emph{volume preserving gauge}. In other words,
the (spatial) average of the inhomogeneous geometry in the volume
preserving gauge leads to a geometry with homogeneous and
isotropic spatial sections, also in a volume preserving gauge.
Note that the gauge in \eqn{spatlim11} for the FLRW spacetime
differs from the standard synchronous and comoving gauge, only by
a redefinition of the time coordinate. The vector field $\bar v^a$
introduced at the beginning of this section and which defines the
FLRW spatial sections, is now given by
\begin{equation}
\bar v^a = \left(\ba^3,0,0,0\right) ~~;~~ \bar v_a = G_{ab}\bar
v^b = \left(-\frac{1}{\ba^3},0,0,0,\right) \,. \label{spatlim12}
\end{equation}
%

Before proceeding to the calculation of the correlation terms and
the averaged Einstein equations, we briefly describe why it is
important to consider the spatial averaging limit of the MG
averaging operation. The key idea to emphasize is that an average
of the homogeneous and isotropic FLRW geometry, should give back
the same geometry. Since the FLRW geometry has a preferred set of
spatial sections, it is important therefore to perform the
averaging over these sections. Further, since the FLRW metric
adapted to its preferred spatial sections depends on the time
coordinate, it is also essential that the spacetime average should
involve a time range that is short compared to the scale over
which say the scale factor changes significantly. (See also Sec. 4 of
\Cite{mars}.) Clearly then, averaging the FLRW metric (denoted
$^{(FLRW)}g_{ab}$) given in \eqn{spatlim11} (which is in volume
preserving gauge) will strictly yield the same metric \emph{only} in
the limit $T\to0$. Namely, for the cuboid $\bm{\Sigma}$ defined in
\eqn{spatlim6} 
\begin{align}
\avg{^{(FLRW)}\bil{g}_{ab}} &=
\lim_{T\to0}\frac{1}{TL^3}\int_\Sigma{dt^\prime d^3x^\prime\,
  ^{(FLRW)}g_{ab}(t^\prime,\rmb{x}^\prime)} \nonumber\\ 
&=\,^{(FLRW)}g_{ab}\,, \label{spatlim14}
\end{align}
which should be clear from the definition of the metric. The
result $\avg{^{(FLRW)}\bil{g}_{ab}} = \,^{(FLRW)}g_{ab}$ in the
spatial averaging limit can also be shown to hold for the FLRW
metric in synchronous gauge, where the coordination bivector
$\W{a}{j}$ can be easily computed using the transformation from
the VPCs $(\bt,x^A)$ to the synchronous coordinates $(\tau,y^A)$
given by
\begin{equation}
\tau = \int^{\bt}{\frac{dt}{\ba^3(t)}} ~~;~~ y^A = x^A\,.
\label{spatlim15}
\end{equation}
The transformation \eqref{spatlim15} will also later allow us to
write the averaged equations in the synchronous gauge for the
averaged geometry.

We now proceed to calculating the correlation $2$-form
$\bZ{a}{b}{i}{j}$ and thereby the averaged Einstein equations.
\label{spatlim}

\section{The correlation $2$-form and the averaged field equations}
\label{correln}
\subsection{Results for the Volume Preserving Gauge}
\noindent We start by defining (in any gauge with $N^A=0$) the
expansion tensor $\Theta{}^A_B$ by
\begin{equation}
\Theta{}^A_B\equiv \frac{1}{2N}h^{AC}\dot h_{CB}\,,
\label{correln1}
\end{equation}
where the dot will always refer to a derivative with respect to
the VPC time $\bt$, and $h^{AB}$ is the inverse of the $3$-metric
$h_{AB}$. (This also gives the symmetric tensor $\Theta_{AB} =
(1/2N)\dot h_{AB}$, which is the negative of the extrinsic
curvature tensor.) The traceless symmetric shear tensor $\sigma{}^A_B$
and the shear scalar $\sigma^2$ are defined by
\begin{equation}
\sigma{}^A_B\equiv \Theta{}^A_B-(\Theta/3) \delta{}^A_B ~~;~~
\sigma^2 \equiv \frac{1}{2}\sigma{}^A_B\sigma{}^B_A\,,
\label{correln2}
\end{equation}
where $\Theta\equiv\Theta{}^A_A = (1/N)\partial_{\bt}\ln{\sqrt h}$ is
the expansion scalar. 

The connection $1$-forms $\om^i_{\ph{i}j} = \Gamma{}^i_{jk}\dx{k}$
can be easily calculated in terms of the expansion tensor, for an
arbitrary lapse function $N$. Specializing to the volume
preserving gauge ($N=h^{-1/2}$), the bilocal extensions
$\Om{i}{j}$ of the connection $1$-forms are trivial and are simply
given by
\begin{equation}
\Om{i}{j}(x^\prime,x) = \Gamma{}^i_{jk}(x^\prime)\dx{k}\,.
\label{correln6}
\end{equation}
Since $G_{ab} = \bar g_{ab}$, the connection $1$-forms
$\barOm{i}{j}$ for the averaged manifold \Mbar\ are constructed
using the FLRW metric in volume preserving gauge given in
\eqn{spatlim11}, and can also be easily evaluated.

We can now construct the correlation $2$-form $\bZ{a}{b}{i}{j}$
defined in \eqn{recap24}. For completeness, we will display all
the nontrivial components $\bZ{a}{b}{i}{j}$, although not all of
them will be relevant for the final equations. The condition
$N=h^{-1/2}$ has the effect that several of the Christoffel
symbols become related to each other. For example, we have
$\Gamma{}^0_{00}=-\partial_{\bt}(\ln\sqrt{h}) = -\Gamma{}^A_{0A} =
-(1/\sqrt{h})\Theta$, and so on. In the following,
$^{(3)}\Gamma{}^A_{BC}$ denotes the Christoffel symbol built from
the $3$-metric $h_{AB}$, and we have defined
\begin{equation}
H\equiv \frac{1}{\ba}\frac{d\ba}{d\bt}\,. \label{correln8}
\end{equation}
\begin{widetext}
We have,
\begin{subequations}
\begin{align}
\bZ{0}{0}{0}{A}&= - \bZ{0}{A}{0}{0} \nonumber\\
               &= \left[\avg{\Theta\Theta_{AJ}} +
               \avg{\partial_A(\ln\sqrt{h})\partial_J(\ln\sqrt{h})} -
               3\ba^8H^2\delta_{AJ}\right] \dx{J} \wedge \ext\bt +
               \avg{\sqrt{h}\Theta_{AJ}\partial_K(\ln\sqrt{h})}\dx{J}
               \wedge \dx{K}    \,,
\label{correln9a} \\&\nonumber\\
\bZ{0}{0}{A}{0}&= - \bZ{A}{0}{0}{0} \nonumber\\
               &= \left[ \avg{\frac{1}{h}\Theta\Theta{}^A_J} +
               \avg{\frac{1}{h}h^{AK} \partial_K(\ln\sqrt{h})
               \partial_J(\ln\sqrt{h})} - 3H^2\delta{}^A_J
               \right]\dx{J} \wedge \ext\bt  +  \avg{\frac{1}{\sqrt
               h}\Theta{}^A_J \partial_K(\ln\sqrt{h})} \dx{J} \wedge
               \dx{K} \,,
\label{correln9b} \\&\nonumber\\
\bZ{0}{0}{A}{B}&= - \bZ{A}{B}{0}{0} \nonumber\\
               &= \left[
               \avg{\frac{1}{\sqrt{h}}\Theta\,^{(3)}\Gamma{}^A_{BJ}} -
               \avg{\frac{1}{\sqrt h} \Theta{}^A_B
               \partial_J(\ln\sqrt{h})} \right] \dx{J} \wedge \ext\bt
               + \avg{\,^{(3)}\Gamma{}^A_{BJ}\partial_K(\ln\sqrt{h})}
               \dx{J} \wedge \dx{K} \,,
\label{correln9c} \\&\nonumber\\
\bZ{0}{A}{0}{B}&= \left[2
  \avg{\sqrt{h}\partial_{\left[A\right.}(\ln\sqrt{h})
  \Theta_{\left.B\right]J}} \right] \dx{J}  \wedge \ext\bt + \left[
  \avg{h\Theta_{AJ}\Theta_{BK}} -  \ba^{16}H^2\delta_{AJ}\delta_{BK}
  \right] \dx{J} \wedge \dx{K} \,,
\label{correln9d} \\&\nonumber\\
\bZ{0}{A}{B}{0}&= - \bZ{B}{0}{0}{A} \nonumber\\
               &= \left[ \avg{\frac{1}{\sqrt
               h}\partial_A(\ln\sqrt{h})\Theta{}^B_J} -
               \avg{\frac{1}{\sqrt  h}h^{BK}\partial_K(\ln\sqrt{h})
               \Theta_{AJ}} \right] \dx{J} \wedge \ext\bt  + \left[
               \avg{\Theta_{AJ}\Theta{}^B_K} -
               \ba^8H^2\delta_{AJ}\delta{}^B_K\right]
               \dx{J}\wedge\dx{K}   \,,
\label{correln9e} \\&\nonumber\\
\bZ{0}{A}{B}{C}&= - \bZ{B}{C}{0}{A} \nonumber\\
               &= \left[
               \avg{\partial_A(\ln\sqrt{h})\,^{(3)}\Gamma{}^B_{CJ}} +
               \avg{\Theta_{AJ}\Theta{}^B_C} -
               \ba^8H^2\delta_{AJ}\delta{}^B_C \right] \dx{J} \wedge
               \ext\bt +
               \avg{\sqrt{h}\Theta_{AJ}\,^{(3)}\Gamma{}^B_{CK}} \dx{J}
               \wedge \dx{K} \,,
\label{correln9f} \\&\nonumber\\
\bZ{A}{0}{B}{0}&= \left[2
  \avg{\frac{1}{h^{3/2}}\partial_K(\ln\sqrt{h})h^{K\left[A\right.}
  \Theta{}^{\left.B\right]}_J} \right] \dx{J} \wedge \ext\bt + \left[
  \avg{\frac{1}{h}\Theta{}^A_J\Theta{}^B_K} -
  H^2\delta{}^A_J\delta{}^B_K \right] \dx{J} \wedge \dx{K} \,,
\label{correln9g} \\&\nonumber\\
\bZ{A}{0}{B}{C}&= - \bZ{B}{C}{A}{0} \nonumber\\
               &= \left[
               \avg{\frac{1}{h}h^{AK}\partial_K(\ln\sqrt{h})
               \,^{(3)}\Gamma{}^B_{CJ}} +
               \avg{\frac{1}{h}\Theta{}^A_J\Theta{}^B_C} -
               H^2\delta{}^A_J\delta{}^B_C \right] \dx{J} \wedge
               \ext\bt + \avg{\frac{1}{\sqrt
               h}\Theta{}^A_J\,^{(3)}\Gamma{}^B_{CK}} \dx{J} \wedge
               \dx{K}  \,,
\label{correln9h} \\&\nonumber\\
\bZ{A}{B}{C}{D} &= \left[\avg{\frac{1}{\sqrt h} \Theta{}^C_D
    \,^{(3)}\Gamma{}^A_{BJ}} - \avg{\frac{1}{\sqrt h} \Theta{}^A_B
    \,^{(3)}\Gamma{}^C_{DJ}}\right] \dx{J} \wedge \ext\bt +
    \avg{\,^{(3)}\Gamma{}^A_{BJ}\,^{(3)}\Gamma{}^C_{DK}} \dx{J} \wedge
    \dx{K} \,,
\label{correln9i}
\end{align}
\label{correln9}
\end{subequations}
\end{widetext}
\noindent where we have used the relation $\,^{(3)}\Gamma{}^J_{BJ}
=\partial_B(\ln\sqrt{h})$, and the square brackets around indices
indicate antisymmetrization ($p_{\left[ij\right]} = (1/2!)(p_{ij}
- p_{ji})$. )

It is now straightforward to use the relations in \eqn{recap37}
(note the unconventional normalization of the $2$-form) to read
off the components $Z^{a\ph{bm}i}_{\ph{a}bm\ph{i}jn}$ and hence
perform the required summations to construct $Z^a_{\ph{a}ijb}$.
This, together with the relation $\bar g^{ab} = G^{ab}$, allows us
to construct the correlation tensor $C{}^a_b$ defined in
\eqn{recap38}
\begin{equation}
C{}^a_b = \left(Z^a_{\ph{a}ijb} - \frac{1}{2}\delta{}^a_b
Z^m_{\ph{a}ijm} \right)G^{ij}. \label{correln10}
\end{equation}
Now, the components of the Einstein tensor $E{}^a_b$ for the
averaged spacetime with metric \eqref{spatlim11} are given by
\begin{align}
E{}^{\bt}_{\bt} &= 3\ba^6H^2 ~~;~~ E{}^{\bt}_A = 0 = E{}^B_{\bt}
\,,
\nonumber\\
E{}^A_B &= \ba^6\delta{}^A_B\left[ 2\left(\frac{\ddot\ba}{\ba} +
  3H^2\right) + H^2\right] \,,
\label{correln11}
\end{align}
where the peculiar splitting of terms in the last equation is for
later convenience. Recall that the overdot denotes a derivative
with respect to the VPC time $\bt$, not synchronous time. In terms
of the coordinate independent objects introduced in
\eqn{spatlim-FLRW1}, we have
\begin{equation}
j_1(x) = -3\ba^6H^2 ~~;~~ j_2(x) = \ba^6\left[
  2\left(\frac{\ddot\ba}{\ba} + 3H^2\right) + H^2\right] \,.
\label{correln-FLRW1}
\end{equation}
>From the averaged Einstein equations in (\ref{recap40}) we next
construct the scalar equations which in the standard case would
correspond to the Friedmann equation and the Raychaudhuri
equation. These correspond to the Einstein tensor components,
\begin{equation}
E{}^a_b\bar v^b\bar v_a  = j_1(x)  ~~;~~ \pi^b_aE{}^a_b +
E{}^a_b\bar v^b\bar v_a = 3j_2(x) + j_1(x)\,, \label{correln12}
\end{equation}
and are given by
\begin{subequations}
\begin{align}
&3\ba^6H^2 =\left(\kappa T{}^a_b - C{}^a_b\right)\bar v_a\bar v^b
  \nonumber\\ 
&~~~~~~~~~=\kappa\bar\rho - \frac{1}{2}\left[ \Cal{Q}^{(1)} +
  \Cal{S}^{(1)} \right]  \,,
\label{correln13a} \\&\nonumber\\
&6\ba^6\left( \frac{\ddot\ba}{\ba} + 3H^2 \right) = \left(-\kappa 
T{}^a_b + C{}^a_b\right) \left(\bar v_a\bar v^b + \pi^b_a \right)
\nonumber\\ 
&~~~~~~~~~~= -\kappa\left(\bar\rho+3\bar p\right) + 2\left[
  \Cal{Q}^{(1)} +  \Cal{Q}^{(2)} + \Cal{S}^{(2)} \right]\,.
\label{correln13b}
\end{align}
\label{correln13}
\end{subequations}
\noindent Here \eqn{correln13a} is the modified Friedmann equation
and \eqn{correln13b} the modified Raychaudhuri equation (in the
volume preserving gauge on \Mbar). We have used
\eqn{spatlim-T-ab1}, with the overbar on $\rho$ and $p$ reminding
us that they are expressed in terms of the nonsynchronous time
$\bt$, and we have defined the correlation terms
\begin{subequations}
\begin{align}
\Cal{Q}^{(1)} &=
\ba^6\left[\frac{2}{3}\left(\avg{\frac{1}{h}\Theta^2} -
\frac{1}{\ba^6}(^{\rm F}\Theta^2)\right) -
2\avg{\frac{1}{h}\sigma^2}\right] ~; \nonumber\\
&~~~~~~~~~~~~~~~~~~~~~~~~~~~~~~~~ \frac{1}{\ba^6}(^{\rm F}\Theta^2)
= \left(3H\right)^2\,, 
\label{correln14a} \\&\nonumber\\
\Cal{S}^{(1)} &= \frac{1}{\ba^2}\delta^{AB}\left[
  \avg{\,^{(3)}\Gamma{}^J_{AC}\,^{(3)}\Gamma{}^C_{BJ}}
  \right. \nonumber\\ 
&~~~~~~~~~~~~~~~~~~\left. -
  \avg{\partial_A(\ln\sqrt{h})\partial_B(\ln\sqrt{h})}
   \right] \,,
\label{correln14b} \\&\nonumber\\
\Cal{Q}^{(2)} &= \ba^6\avg{\frac{1}{h}\Theta{}^A_B\Theta{}^B_A} -
\frac{1}{\ba^2}\delta^{AB}\avg{\Theta_{AJ}\Theta{}^J_B},
\label{correln14c} \\&\nonumber\\
\Cal{S}^{(2)} &= \ba^6\avg{\frac{1}{h}h^{AB}
  \partial_A(\ln\sqrt{h})\partial_B(\ln\sqrt{h})} \nonumber\\ 
&~~~~~~~~~~~-  \frac{1}{\ba^2}\delta^{AB}
  \avg{\partial_A(\ln\sqrt{h})\partial_B(\ln\sqrt{h})}  \,.
\label{correln14d}
\end{align}
\label{correln14}
\end{subequations}
%
In defining $\Cal{Q}^{(1)}$ we have used the relation $\Theta^2 -
\Theta{}^A_B\Theta{}^B_A = (2/3)\Theta^2 - 2\sigma^2$.
$\Cal{Q}^{(1)}$ and $\Cal{Q}^{(2)}$ are correlations of the
extrinsic curvature, whereas $\Cal{S}^{(1)}$ and $\Cal{S}^{(2)}$
are correlations restricted to the intrinsic $3$-geometry of the
spatial slices of \Cal{M}. Since the components of $C{}^a_b$ are
not explicitly constrained by \eqns{recap29} and
\eqref{recap29add}, we can treat the combinations
$(1/2)(\Cal{Q}^{(1)} + \Cal{S}^{(1)})=-C{}^0_0$ and
$2(\Cal{Q}^{(1)} + \Cal{Q}^{(2)} + \Cal{S}^{(2)}) =
(C{}^A_A-C{}^0_0)$ as independent, subject only to the
differential constraints \eqref{recap42} which we will come to
below.

As discussed in the beginning of Section \ref{spatlim}, the
remaining components of $C{}^a_b$ must be set to zero, giving
constraints on the underlying inhomogeneous geometry. In
coordinate independent language, these constraints read
\begin{align}
\pi^b_k C{}^a_b\bar v_a &= 0 =\,\pi^k_a C{}^a_b\bar v^b ~;\nonumber\\ 
\pi^i_a\pi^b_k C{}^a_b &- \frac{1}{3}\pi^i_k \left(\pi^b_a
C{}^a_b\right) = 0\,. \label{correln15}
\end{align}
\eqns{correln15} reduce to the following for our specific choice
of volume preserving coordinates,
\begin{equation}
C{}^0_A = 0 ~~;~~ C{}^A_0 = 0 ~~;~~ C{}^A_B -
\frac{1}{3}\delta{}^A_B(C{}^J_J) = 0 \,, \label{correln16}
\end{equation}
It can be shown that the VPC assumption $N=h^{-1/2}$ reduces the
correlations $\Cal{Q}^{(2)}$ and $\Cal{S}^{(2)}$ defined in
\eqns{correln14c} and \eqref{correln14d}, as well as several terms
in the explicit expansion of \eqn{correln16}, to the form
\begin{equation}
\frac{1}{\avg{g_{00}}}\avg{g_{00}g^{AB}\Gamma{}^{a_1}_{b_1c_1}
  \Gamma{}^{i_1}_{j_1k_1}} - \avg{g^{AB}}\avg{\Gamma{}^{a_2}_{b_2c_2}
  \Gamma{}^{i_2}_{j_2k_2}}\,.
\label{correln18}
\end{equation}
The assumption in \eqn{recap31b} (which is fundamental to the MG
formalism) now shows that one can write
\begin{align}
\avg{g_{00}g^{AB}\Gamma{}^a_{bc}\Gamma{}^i_{jk}} &=
\avg{g_{00}g^{AB}}\avg{\Gamma{}^a_{bc}\Gamma{}^i_{jk}} \nonumber\\ 
&= -\avg{\frac{h^{AB}}{h}}\avg{\Gamma{}^a_{bc}\Gamma{}^i_{jk}} \,.
\label{correln19}
\end{align}
An interesting point is that the VPC assumption $N=h^{-1/2}$
further allows us to assume $\avg{h^{AB}/h} =
\avg{h^{AB}}\avg{1/h}$ consistently with the formalism (details in
Appendix \eqref{app-gamma}). Using \eqn{spatlim10} this gives us
\begin{equation}
\avg{\frac{h^{AB}}{h}} = \frac{1}{\ba^6}\avg{h^{AB}}\,.
\label{correln20}
\end{equation}
This  shows that the correlation terms $\Cal{Q}^{(2)}$ and
$\Cal{S}^{(2)}$ in fact vanish,
\begin{equation}
\Cal{Q}^{(2)} = 0 = \Cal{S}^{(2)}\,, \label{correln21}
\end{equation}
and leads to some remarkable cancellations in \eqns{correln16},
which simplify to give
\begin{subequations}
\begin{align}
&\delta^{JK}\left[\avg{\sqrt{h}\Theta_{JB}\,^{(3)}\Gamma{}^B_{AK}}
-
  \avg{\sqrt{h}\Theta_{JK}\,^{(3)}\Gamma{}^B_{AB}} \right] = 0\,,
\label{correln22a}\\
&\delta^{JK}\avg{\frac{1}{\sqrt
h}\Theta{}^B_K\,^{(3)}\Gamma{}^A_{JB}}
  - \delta^{AJ}\avg{\frac{1}{\sqrt
      h}\Theta{}^K_K\,^{(3)}\Gamma{}^B_{JB}} = 0\,,
\label{correln22b}\\
&\delta^{JK}\avg{\,^{(3)}\Gamma{}^A_{JC}\,^{(3)}\Gamma{}^C_{KB}} -
\delta^{AJ}\avg{\,^{(3)}\Gamma{}^C_{JC}\,^{(3)}\Gamma{}^K_{BK}}
\nonumber\\ 
&~~~~~~~~~~~~~~~~~~~~~~~~~~~~~~~~~~~~~~~~~~~~~=
\frac{1}{3}\delta{}^A_B\left(\ba^2\Cal{S}^{(1)}\right) \,.
\label{correln22c}
\end{align}
\label{correln22}
\end{subequations}
These simplifications are solely a consequence of assuming that
the inhomogeneous metric in the volume preserving gauge averages
out to give the FLRW metric in standard form. In general, these
simplifications will not occur when the standard FLRW metric
arises from an arbitrary choice of gauge for the inhomogeneous
metric.

In order to come as close as possible to the standard approach in
Cosmology, we will now rewrite the scalar equations
\eqref{correln13} (which are the cosmologically relevant ones)
after performing the transformation given in \eqn{spatlim15} in
order to get the FLRW metric to the form
\begin{equation}
^{(\Mbar)}ds^2 = -d\tau^2 + a^2(\tau)\delta_{AB}dy^Ady^B ~~;~~
a(\tau)
  =  \ba(\bt(\tau))\,.
\label{correln23}
\end{equation}
Since \eqns{correln13} are \emph{scalar} equations, this
transformation only has the effect of reexpressing all the terms
as functions of the synchronous time $\tau$. Although the
transformation will change the explicit form of the coordination
bivector $\W{a}{j}$, this change involves only the time
coordinate, and in the spatial averaging limit there is no
difference between averages computed in the VPCs and those
computed after the time redefinition. This again emphasizes the
importance of the spatial averaging limit of spacetime averaging,
if we are to succeed operationally in explicitly displaying the
correlations as corrections to the standard cosmological
equations. The correlation terms in \eqns{correln14} are therefore
still interpreted with respect to the volume preserving gauge, but
are treated as functions of $\tau$. For the scale factor on the
other hand, we have
\begin{equation}
\ba^3H = \frac{1}{a}\frac{da}{d\tau} \equiv H_{\rm FLRW} ~~;~~
\ba^6\left(\frac{\ddot\ba}{\ba}+3H^2\right) =
\frac{1}{a}\frac{d^2a}{d\tau^2} \,. \label{correln24}
\end{equation}
Further writing
\begin{equation}
\rho(\tau) = \bar\rho(\bt(\tau))  ~~;~~ p(\tau) = \bar
p(\bt(\tau)) \,, \label{correln25}
\end{equation}
equations \eqref{correln13} become
\begin{subequations}
\begin{align}
H^2_{\rm FLRW} &= \frac{8\pi G_N}{3}\rho -  \frac{1}{6}\left[
  \Cal{Q}^{(1)} + \Cal{S}^{(1)} \right],
\label{correln26a} \\&\nonumber\\
\frac{1}{a}\frac{d^2a}{d\tau^2} &= -\frac{4\pi G_N}{3}\left(\rho +
  3p\right) + \frac{1}{3}\Cal{Q}^{(1)} \,.
\label{correln26b}
\end{align}
\label{correln26}
\end{subequations}
We emphasize that the quantities $\Cal{Q}^{(1)}$ and
$\Cal{S}^{(1)}$, defined in \eqns{correln14a} and
\eqref{correln14b} as correlations in the \emph{volume preserving}
gauge, are to be thought of as functions of the \emph{synchronous}
time $\tau$, where the coordinate $\tau$ itself was defined
\emph{after} the spatial averaging. Such an identification is
justified since we are dealing with scalar combinations of these
quantities. Note that $\Cal{Q}^{(1)}$ and $\Cal{S}^{(1)}$ can be
treated independently, apart from the constraints imposed by
\eqn{recap42}, which we turn to next. These conservation
conditions can be decomposed into a scalar part and a $3$-vector
part, given respectively by
\begin{equation}
\bar v^bC{}^a_{b;a} = 0 ~~;~~ \pi^b_k C{}^a_{b;a} = 0 \,.
\label{correln27}
\end{equation}
In the synchronous gauge \eqref{correln23} for the FLRW metric,
the scalar equation reads
\begin{equation}
\left(\partial_\tau\Cal{Q}^{(1)} + 6H_{\rm
FLRW}\Cal{Q}^{(1)}\right) + \left(\partial_\tau\Cal{S}^{(1)} +
2H_{\rm FLRW}\Cal{S}^{(1)}\right) = 0 \,. \label{correln28}
\end{equation}
We recall that this equation is a consequence of setting the
correlation 3-form and the correlation 4-form to zero, and it
relates the evolution of $\Cal{Q}^{(1)}$ and $\Cal{S}^{(1)}$. The
$3$-vector equation (on imposing the first set of conditions in
\eqn{correln15}) simply gives $\partial_\tau C{}^\tau_A = 0$, so
that $C{}^\tau_A = 0 =\,$constant, which also implies that
$C{}^A_\tau = 0=\,$constant and hence this equation gives nothing
new. (We have used the relations $C{}^0_0 = C{}^\tau_\tau$,
$C{}^0_A = \ba^3 C{}^\tau_A$ and $C{}^A_0 = (1/\ba^3)C{}^A_\tau$
where $0$ denotes the nonsynchronous time coordinate $\bt$.)

The cosmological equations (\ref{correln26}), along with the
constraint equations (\ref{correln22}) and (\ref{correln28}) are
the key results of this section. Subject to the acceptance of the
volume preserving gauge on the underlying manifold ${\cal M}$ they
can in principle be used to study the role of the correction terms
resulting from spatial averaging.

\subsection{Results for an arbitrary gauge choice}
\label{gauge} \noindent In this subsection, we will display the
results obtained on assuming that the metric
\begin{equation}
^{(\Cal{M})}ds^2 = - N^2(t,\rmb{x})dt^2  +
  h_{AB}(t,\rmb{x})dx^Adx^B\,,
\label{gauge1}
\end{equation}
averages out to the FLRW metric in standard form with a
nonsynchronous time coordinate $t$ in general, to give
\begin{equation}
^{(\Mbar)}ds^2 = -f^2(t)dt^2 + \ba^2(t)\delta_{AB}dx^Adx^B\,.
\label{gauge2}
\end{equation}
In other words, we are assuming that the relations in
\eqn{spatlim-gauge1} hold. Note that the averaging operation is no
longer trivial, although we are still assuming an averaging on
domains corresponding to ``thin'' time slices. We again split the
averaged Einstein equations into scalar equations, and $3$-vector
and traceless $3$-tensor equations. After transforming to the
synchronous time coordinate $\tau$, now defined by
\begin{equation}
\tau = \int^{t}{f(t^\prime)dt^\prime}\,, \label{gauge3}
\end{equation}
and again defining $H\equiv(1/\ba)(d\ba/dt)$ and $H_{\rm
  FLRW}\equiv(1/a)(da/d\tau)$ with $a(\tau) = \ba(t(\tau))$, the
modified Friedmann and Raychaudhuri equations read
\begin{subequations}
\begin{align}
&H^2_{\rm FLRW} = \frac{8\pi G_N}{3}\rho -  \frac{1}{6}\left[
  \tilde{\Cal{P}}^{(1)} + \tilde{\Cal{S}}^{(1)} \right]  \,,
\label{gauge4a}\\ &\nonumber\\
&\frac{1}{a}\frac{d^2a}{d\tau^2} = -\frac{4\pi
  G_N}{3}\left(\rho+3p\right) + \frac{1}{3}\left[
  \tilde{\Cal{P}}^{(1)} + \tilde{\Cal{P}}^{(2)} +
  \tilde{\Cal{S}}^{(2)} \right] \,,
\label{gauge4b}
\end{align}
\label{gauge4}
\end{subequations}
where the correlation terms are now defined using the relations,
\begin{subequations}
\begin{align}
&\tilde{\Cal{P}}^{(1)} = \frac{1}{f^2}\left[
  \avg{\bil{\Gamma}{}^A_{0A}\bil{\Gamma}{}^B_{0B}} -
  \avg{\bil{\Gamma}{}^A_{0B}\bil{\Gamma}{}^B_{0A}} - 6H^2\right] \,,
\label{gauge5a} \\&\nonumber\\
&\tilde{\Cal{S}}^{(1)} =
\avg{\bil{g}^{JK}}\left[\avg{\bil{\Gamma}{}^A_{JB}\bil{\Gamma}{}^B_{KA}}
  - \avg{\bil{\Gamma}{}^A_{JA}\bil{\Gamma}{}^B_{KB}}\right] \,,
\label{gauge5b} \\&\nonumber\\
&\tilde{\Cal{P}}^{(2)} + \tilde{\Cal{P}}^{(1)} =
-\frac{1}{f^2}\avg{\bil{\Gamma}{}^A_{0A}\bil{\Gamma}{}^0_{00}} -
\avg{\bil{g}^{JK}}\avg{\bil{\Gamma}{}^0_{JA}\bil{\Gamma}{}^A_{0K}}
\nonumber\\ 
&~~~~~~~~~~~~~~~~~~+
\frac{3H}{f^2}\left(\partial_{t}(\ln f) + H\right) \,, 
\label{gauge5c} \\&\nonumber\\
&\tilde{\Cal{S}}^{(2)} =
\frac{1}{f^2}\avg{\bil{\Gamma}{}^A_{00}\bil{\Gamma}{}^0_{A0}} +
\avg{\bil{g}{}^{JK}}\avg{\bil{\Gamma}{}^0_{J0}\bil{\Gamma}{}^A_{KA}}
\,. \label{gauge5d}
\end{align}
\label{gauge5}
\end{subequations}
We emphasize that averaging here refers to spatial averaging. Also
$\avg{\bil{g}^{JK}} = G^{JK} = (1/\ba^2)\delta^{JK}$, and the
index $0$ refers to the nonsynchronous time $t$. It is easy to
check that $\tilde{\Cal{P}}^{(1)}$ and $\tilde{\Cal{P}}^{(1)} +
\tilde{\Cal{P}}^{(2)}$ correspond to correlations of (the bilocal
extensions of) the extrinsic curvature with itself and with the
time derivative of the lapse function. $\tilde{\Cal{S}}^{(1)}$
corresponds to correlations between the bilocal extensions of the
Christoffel symbols of the $3$-geometry, and
$\tilde{\Cal{S}}^{(2)}$ to correlations of the extension of the
spatial derivative of the lapse function with itself and with the
Christoffel symbols of the $3$-geometry. Due to the way we have
defined these correlations, one can also check that when the lapse
function satisfies $N\sqrt{h}=1$ (so that the averaging becomes
trivial), we have $\tilde{\Cal{P}}^{(1)} = {\Cal{Q}}^{(1)}$,
$\tilde{\Cal{S}}^{(1)}={\Cal{S}}^{(1)}$, and
$\tilde{\Cal{P}}^{(2)} = 0 = \tilde{\Cal{S}}^{(2)}$, where
${\Cal{Q}}^{(1)}$ and ${\Cal{S}}^{(1)}$ were defined in
\eqns{correln14}. The $3$-vector and traceless $3$-tensor
equations become
\begin{subequations}
\begin{align}
&\frac{1}{f^2}\left[\avg{\bil{\Gamma}{}^0_{0A}\bil{\Gamma}{}^B_{B0}}
- 
  \avg{\bil{\Gamma}{}^0_{0B}\bil{\Gamma}{}^B_{A0}}\right] \nonumber\\
&\ph{\frac{1}{f^2}[\avg{\bil{\Gamma}{}^0_{0A}\bil{\Gamma}{}^B_{B0}}-]}
+  \avg{\bil{g}^{JK}}\left[
  \avg{\bil{\Gamma}{}^0_{JB}\bil{\Gamma}{}^B_{AK}}
  - \avg{\bil{\Gamma}{}^0_{JA}\bil{\Gamma}{}^B_{BK}} \right] = 0\,
\label{gauge6a} \\&\nonumber\\
&\frac{1}{f^2}\left[\avg{\bil{\Gamma}{}^A_{00}\bil{\Gamma}{}^B_{B0}}
-
  \avg{\bil{\Gamma}{}^B_{00}\bil{\Gamma}{}^A_{B0}}\right] \nonumber\\
&\ph{\frac{1}{f^2}[\avg{\bil{\Gamma}{}^0_{0A}\bil{\Gamma}{}^B_{B0}}-]}
  +  \avg{\bil{g}^{JK}}\left[
  \avg{\bil{\Gamma}{}^A_{JB}\bil{\Gamma}{}^B_{0K}}
  - \avg{\bil{\Gamma}{}^A_{J0}\bil{\Gamma}{}^B_{BK}} \right] = 0\,
\label{gauge6b}\\&\nonumber\\
&\frac{1}{f^2}\left[\avg{\bil{\Gamma}{}^A_{B0}\bil{\Gamma}{}^m_{0m}}
-
 \avg{\bil{\Gamma}{}^A_{m0}\bil{\Gamma}{}^m_{0B}} \right] \nonumber\\
&~~~+ \avg{\bil{g}^{JK}}
 \left[\avg{\bil{\Gamma}{}^A_{Jm}\bil{\Gamma}{}^m_{KB}} -
 \avg{\bil{\Gamma}{}^A_{JB}\bil{\Gamma}{}^m_{Km}} \right]
 \nonumber\\
&~~~=
 -\frac{1}{3}\delta{}^A_B\left[\tilde{\Cal{P}}^{(2)} +
 \tilde{\Cal{S}}^{(2)} - \tilde{\Cal{S}}^{(1)} - \frac{9H}{f^2}\left(
 H + \frac{1}{3}\partial_{t}(\ln f) \right) \right] \,,
\label{gauge6c}
\end{align}
\label{gauge6}
\end{subequations}
where the lower case index $m$ in the last equation runs over all
spacetime indices $0,1,2,3$, with the index $0$ referring to the
nonsynchronous time $t$. It is easy to check that \eqns{gauge6}
reduce to \eqns{correln22} with the choice $N=h^{-1/2}$. The
condition $C{}^a_{b;a}=0$ has the scalar part,
\begin{align}
\left(\partial_\tau\tilde{\Cal{P}}^{(1)} + 6H_{\rm
  FLRW}\tilde{\Cal{P}}^{(1)}\right) &+
  \left(\partial_\tau\tilde{\Cal{S}}^{(1)} + 2H_{\rm
  FLRW}\tilde{\Cal{S}}^{(1)}\right) \nonumber\\
& + 4H_{\rm FLRW}\left(\tilde{\Cal{P}}^{(2)} + \tilde{\Cal{S}}^{(2)}
  \right) = 0 
  \,,
\label{gauge7}
\end{align}
while the $3$-vector part, as before, gives nothing new and simply
states $\partial_\tau C{}^\tau_A=0$.

We can now state the main result of our paper in a clear and
unambiguous manner, as follows : Having assumed that the FLRW
spatial sections arise as the average of some gauge choice with
lapse function $N(t,\rmb{x})$, spatial $3$-metric
$h_{AB}(t,\rmb{x})$ and shift vector $N^A$ set to zero for
convenience, we can construct the \emph{scalar} quantities
$C{}^a_b\bar v^b\bar v_a$ and $\pi^b_a C{}^a_b + C{}^a_b\bar
v^b\bar v_a$ which, in coordinates natural to the FLRW metric take
the form,
\begin{align}
&C{}^a_b\bar v^b\bar v_a = \frac{1}{2}\left[\tilde{\Cal{P}}^{(1)} +
  \tilde{\Cal{S}}^{(1)} \right] ~; \nonumber\\
& \pi^b_a C{}^a_b + C{}^a_b\bar
  v^b\bar v_a = 2\left[\tilde{\Cal{P}}^{(1)} + \tilde{\Cal{P}}^{(2)} +
  \tilde{\Cal{S}}^{(2)}\right] \,,
\label{gauge8}
\end{align}
with the various quantities being defined in \eqns{gauge5}. These
scalars modify the usual cosmological equations as shown in
\eqns{gauge4}, and are themselves subject to the differential
conditions \eqref{gauge7}. In addition, for consistency of our
assumptions with the formalism, the underlying inhomogeneous
metric is also subject to the conditions \eqref{gauge6}.

The combinations on the right hand sides of the relations
\eqref{gauge8} can clearly be treated independently, apart from
the conditions \eqref{gauge7}. Further, since the correlation
$2$-form has 40 independent components
$Z^{a\ph{bm}i}_{\ph{a}bm\ph{i}jn}$ after imposing all algebraic
constraints, and since none of the four quantities
$\tilde{\Cal{P}}^{(1)}$, $\tilde{\Cal{P}}^{(2)}$,
$\tilde{\Cal{S}}^{(1)}$ and $\tilde{\Cal{S}}^{(2)}$ are trivially
related by these constraints, one can always treat these four
functions independently of each other, subject only to the
constraint in \eqn{gauge7}. Before proceeding, we wish to make two
remarks concerning the possible behaviour of the correction terms.
It was mentioned in \Cite{coleyPRL} that assuming only spatial
correlations, i.e. assuming that all components of the correlation
$2$-form $\bZ{a}{b}{i}{j}$ with at least one $0$ index vanish, the
corrections must be of the form of a spatial curvature term in the
FLRW equations. We can confirm this statement, since the above
assumption amounts to setting $\tilde{\Cal{P}}^{(1)}$,
$\tilde{\Cal{P}}^{(2)}$ and $\tilde{\Cal{S}}^{(2)}$ to zero,
leaving only $\tilde{\Cal{S}}^{(1)}$ which must then evolve as
$\sim a^{-2}$ because of \eqn{gauge7}. Further, the main result of
the \Cite{coleyPRL} was that assuming the averaged metric to be
FLRW in the \emph{conformal} gauge (which corresponds to assuming
$f(t)=\ba(t)$ in our case) and further assuming all components of
the correlation $2$-form to be \emph{constant}, the corrections
must again be of the form of a spatial curvature term.
\eqns{gauge5} show that this is indeed the case, and \eqn{gauge7}
shows that in this case one must have $\tilde{\Cal{P}}^{(1)} +
\tilde{\Cal{P}}^{(2)} + \tilde{\Cal{S}}^{(2)} = 0$, with no
condition on the constant $a^2\tilde{\Cal{S}}^{(1)}$, which is
consistent with a curvature term in the modified FLRW equations
\eqref{gauge4}. Clearly though, the \emph{allowed} behaviour of
the correction terms which is consistent with \eqn{gauge7}, is 
more general than that of a spatial curvature term, and it is not
yet clear how these corrections behave in the real Universe.

It is only fair to say that much of what we have described is
already implicit in the work of Zalaletdinov and collaborators.
What we have done here is to spell it out explicitly, emphasizing
the relevance of spatial averaging for Cosmology.  Further, the
correction terms resulting from  averaging have been displayed
explicitly as scalars, which could be of help in applications and
comparison with observations. A hitherto unappreciated fact which
emerges is that even if some inhomogeneous geometry yields an FLRW
Universe upon averaging over sufficiently large scales, this is
not sufficient to guarantee consistency with the averaged Einstein
equations. The consistency conditions (\ref{gauge6}) must be
satisfied -- this fact could have potential significance in
restricting the class of allowed initial perturbations in the
early Universe, and should be investigated further.

Another important goal of our paper is to attempt to compare the
approaches of Zalaletdinov and Buchert, a topic to which we now
turn.

\subsection{A comparison with the  averaging formalism of Buchert}
\noindent The averaging formalism developed by Buchert is based
exclusively on the manifold ${\cal M}$, and there is no analog of
the averaged manifold ${\cal \bar{M}}$ in this scheme. Given an
inhomogeneous metric on ${\cal M}$ one takes the trace of the
Einstein equations in the {\it inhomogeneous} geometry, and
carries out a spatial averaging of the inhomogeneous scalar
equations.

We recall in brief Buchert's construction \cite{buchDust}, by
first writing down the averaged equations for the simplest case of
pressureless and irrotational inhomogeneous dust. The metric can
be written in synchronous and comoving gauge as
\begin{equation}
ds^2=\,-dt^2+b_{AB}(\rmb{x},t)dx^Adx^B\,. \label{avg1}
\end{equation}
The Einstein equations can be split \cite{buchDust} into a set of
scalar equations and a set of vector and traceless tensor
equations. The scalar equations are the Hamiltonian constraint
\eqref{avg2a} and the evolution equation for $\Theta$
\eqref{avg2b},
\begin{subequations}
\begin{equation}
\Cal{R}+\frac{2}{3}\Theta^2-2\sigma^2=16\pi G\rho\,, \label{avg2a}
\end{equation}
\begin{equation}
\Cal{R}+\partial_t\Theta+\Theta^2=12\pi G\rho\,, \label{avg2b}
\end{equation}
\label{avg2}
\end{subequations}
where \Cal{R}\ is the Ricci scalar of the 3-dimensional
hypersurface of constant $t$, $\Theta$ and $\sigma^2$ are the
expansion scalar and the shear scalar defined earlier and $\rho$
is the inhomogeneous matter density of the dust. Note that all
quantities in Eqns. \eqref{avg2} generically depend on both
position $\rmb{x}$ and time $t$. Eqns. \eqref{avg2a} and
\eqref{avg2b} can be combined to give Raychaudhuri's equation
\begin{equation}
\partial_t\Theta+\frac{1}{3}\Theta^2+2\sigma^2+4\pi G\rho=0\,.
\label{avg3}
\end{equation}
The continuity equation $\partial_t\rho=-\Theta\rho$ which gives
the evolution of $\rho$, is consistent with Eqns. \eqref{avg2a},
\eqref{avg2b}. Only scalar Einstein equations are considered,
since the spatial average of a scalar quantity can be defined in a
gauge covariant manner, within a given foliation of space-time. We
return to this point below. For the space-time described by
\eqref{avg1}, the spatial average of a scalar $\Psi(\rmb{x},t)$
over a \emph{comoving} domain \Cal{D} at time $t$ is defined by
\begin{equation}
\avgD{\Psi}=\frac{1}{\uD{V}}\int_\Cal{D}{d^3x\sqrt{b}\,\Psi}\,,
\label{avg4}
\end{equation}
where $b$ is the determinant of the 3-metric $b_{AB}$ and $\uD{V}$
is the volume of the comoving domain given by
$\uD{V}=\int_\Cal{D}{d^3x\sqrt{b}}$. Spatial averaging is, by
definition, not generally covariant. Thus the choice of foliation
is relevant, and should be motivated on physical grounds. In the
context of cosmology, averaging over freely-falling observers is a
natural choice, especially when one intends to compare the results
with standard FLRW cosmology.  Following the definition
\eqref{avg4} the following commutation relation then holds
\cite{buchDust}
\begin{equation}
\partial_t\avgD{\Psi}-\avgD{\partial_t\Psi}=
\avgD{\Psi\Theta}-\avgD{\Psi}\avgD{\Theta}\,, \label{avg5}
\end{equation}
which yields for the expansion scalar $\Theta$
\begin{equation}
\partial_t\avgD{\Theta}-\avgD{\partial_t\Theta}=
\avgD{\Theta^2}-\avgD{\Theta}^2\,. \label{avg6}
\end{equation}
Introducing the dimensionless scale factor
$\aD\equiv\left(\uD{V}/V_{\Cal{D} i}\right)^{1/3}$ normalized by
the volume of the domain \Cal{D}\ at some initial time $t_i$, we
can average the scalar Einstein equations \eqref{avg2a},
\eqref{avg2b} and the continuity  equation to obtain
\cite{buchDust}
\begin{subequations}
\begin{equation}
\partial_t\avgD{\rho}=\,-\avgD{\Theta}\avgD{\rho} ~~~;~~~
\avgD{\Theta}=3\frac{\partial_t\aD}{\aD}\,, \label{avg7a}
\end{equation}
\begin{equation}
\left(\frac{\partial_t\aD}{\aD}\right)^2=\frac{8\pi
G}{3}\avgD{\rho} -
\frac{1}{6}\left(\uD{\Cal{Q}}+\avgD{\Cal{R}}\right)\,,
\label{avg7b}
\end{equation}
\begin{equation}
\left(\frac{\partial_t^2\aD}{\aD}\right)= -\frac{4\pi G}{3}
\avgD{\rho}+ \frac{1}{3}\uD{\Cal{Q}}\,. \label{avg7c}
\end{equation}
\label{avg7}
\end{subequations}
Here, the `kinematical backreaction' $\uD{\Cal{Q}}$ is given by
\begin{equation}
\uD{\Cal{Q}}\equiv\frac{2}{3}\left(\avgD{\Theta^2}-
\avgD{\Theta}^2\right)-2\avgD{\sigma^2} \label{avg8}
\end{equation}
and is a spatial constant over the domain \Cal{D}.

A necessary condition for \eqref{avg7c} to integrate to
\eqref{avg7b} takes the form of the following differential
equation involving $\uD{\Cal{Q}}$ and $\avgD{\Cal{R}}$,
\begin{equation}
\partial_t\uD{\Cal{Q}}+6\frac{\partial_t\aD}{\aD}\uD{\Cal{Q}}
+\partial_t\avgD{\Cal{R}}+2\frac{\partial_t\aD}{\aD}\avgD{\Cal{R}}=
0\,. \label{avg9}
\end{equation}

The equations above describe the essence of Buchert's averaging
formalism, for the dust case. We note that the remaining eight
Einstein equations for the inhomogeneous geometry, which are not
scalar equations, are not averaged. These are the five evolution
equations for the trace-free part of the shear,
\begin{equation}
\partial_t\left(\sigma{}^A_B\right) = - \Theta
\sigma{}^A_B -  \Cal{R}{}^A_B +\frac{2}{3}\delta{}^A_B
\left(\sigma^2 - \frac{1}{3}\Theta^2 + 8\pi G\rho \right) \,.
\label{shea1}
\end{equation}
and the three equations relating the spatial variation of the
shear and the expansion,
\begin{equation}
\sigma{}^A_{B || A} = \frac{2}{3} \Theta_{|| B}\,. \label{shea2}
\end{equation}
Here, $\Cal{R}{}^A_B$ is the spatial Ricci tensor and, in
Buchert's notation, a $||$ denotes covariant derivative with
respect to the $3$-metric.

In analogy with the dust case, Buchert's averaging formalism can
be applied to the case of a perfect fluid \cite{buchPress}, by
starting from the metric
\begin{equation}
ds^2 = -N^2 dt^2 + b_{AB}dx^A dx^B \,. \label{Nmet}
\end{equation}
The averaged scalar Einstein equations for the scale factor \aD\
are
\begin{equation}
3\frac{\partial_t^2 \aD}{\aD} + 4\pi G
\avgD{N^2\left(\rho+3p\right)}= \uD{\bar{\Cal{Q}}} +
\uD{\bar{\Cal{P}}} \,, \label{ep1}
\end{equation}
\begin{equation}
6 \uD{H}^2 - 16\pi G \avgD{N^2\rho}= -\uD{\bar{\Cal{Q}}} -
\avgD{N^2\Cal{R}}~~;~~ \uD{H}=\frac{\partial_t\aD}{\aD}\,,
\label{ep2}
\end{equation}
where the kinematical backreaction $\uD{\bar{\Cal{Q}}}$ is given
by
\begin{equation}
\uD{\bar{\Cal{Q}}} = \frac{2}{3}\left(
\avgD{\left(N\Theta\right)^2} - \avgD{N\Theta}^2\right)
 - 2\avgD{N^2\sigma^2}\,,
\label{kb}
\end{equation}
and the dynamical backreaction $\uD{\bar{\Cal{P}}}$ is given by
\begin{equation}
\uD{\bar{\Cal{P}}} = \avgD{N^2\Cal{A}} + \avgD{\Theta\partial_tN}
\,, \label{db}
\end{equation}
where $\Cal{A}=\nabla_j(u^i\nabla_iu^j)$ is the $4$-divergence of
the $4$-acceleration of the fluid. \eqn{ep2} follows as an
integral from \eqn{ep1} if and only if the relation
\begin{align}
\partial_t \uD{\Cal{Q}} &+ 6\uD{H}\uD{\Cal{Q}} +
\partial_t \avgD{N^2\Cal{R}} + 2\uD{H}\avgD{N^2\Cal{R}} +
4\uD{H}\uD{\bar{\Cal{P}}} \nonumber\\
&- 16\pi G \left[\partial_t \avgD{N^2\rho} +
3\uD{H}\avgD{N^2\left(\rho+p\right)} \right]  =0\,, \label{incon}
\end{align}
is satisfied. There are also the unaveraged equations (which we do
not display here) for the shear, analogous to the shear equations
\eqref{shea1} and \eqref{shea2} for dust.

Buchert's approach is the only other approach, apart from
Zalaletdinov's MG, which is capable of treating inhomogeneities in
a nonperturbative manner, although it is limited to using only
scalar quantities within a chosen $3+1$ splitting of spacetime.
Buchert takes the trace of the Einstein equations in the
\emph{inhomogeneous} geometry, and averages these inhomogeneous
scalar equations. In the context of Zalaletdinov's MG however, we
have used the existence of the vector field $\bar v^a$ in the FLRW
spacetime to construct scalar equations \emph{after} averaging the
full Einstein equations. As far as observations are concerned, it
has been noted by Buchert and Carfora \cite{buch-carf}, that the
spatially averaged matter density $\avgD{\rho}$ defined by Buchert
is \emph{not} the appropriate observationally relevant quantity --
the ``observed'' matter density (and pressure) is actually defined
in a \emph{homogeneous} space. Since we have done precisely this
in \eqn{spatlim-T-ab1}, we are directly dealing with the
appropriate observationally relevant quantity in the MG framework.

Another important difference between the two approaches is the
averaging operation itself. Buchert's spatial average, defined for
scalar quantities, is given (for some scalar $\Psi(t,x^A)$) by
(\ref{avg4}) above. On the other hand the averaging operation we
have been using (given by \eqn{spatlim8} using the volume
preserving gauge) is a limit of a spacetime averaging defined
using the coordination bivector $\W{a}{j}$, and is different from
the one in \eqn{avg4}.

Most importantly though, Buchert's averaging scheme by itself does
not incorporate the concept of an averaged manifold \Mbar\
(although the work of Buchert and Carfora \cite{buch-carf} does
deal with $3$-spaces of constant curvature). In a recent paper
\cite{luminous} we had argued that Buchert's ``effective scale
factor''
$a_\Cal{D}(t)\equiv(V_\Cal{D}(t)/V_\Cal{D}(t_{in}))^{1/3}$ must be
the scale factor for the metric of the averaged manifold, upto
some corrections arising due to such effects as calculated by
Buchert and Carfora. In the present work however, it is clear that
such a suggestion is necessarily incomplete due to the presence of
\eqns{correln22} constraining the underlying geometry. These
constraints are in general nontrivial and hence indicate that it
is not sufficient to assume that the metric of the inhomogeneous
manifold averages out to the FLRW form -- there are additional
conditions which the correlations must satisfy.

To our understanding, Buchert's averaging formalism is a valid
aproach, even though it is based on a spatial averaging. A
central difference from the MG approach is the issue of closure :
not all the Einstein equations have been averaged in Buchert's
approach, but only the scalar ones. This puts a constraint on the
allowed solutions considered for the averaged equations:
\eqref{avg7} for the dust case, and \eqref{ep1} and \eqref{ep2}
for the fluid case. Solutions to these equations must necessarily
be checked for consistency with the unaveraged equations for the
shear. Further, averaging over successively larger scales can
bring in additional corrections to the averaged equations, as
discussed by Buchert and Carfora. Also, if one does not wish to
identify Buchert's \aD\ with the scale factor in FLRW cosmology,
one is compelled to develop a whole new set of ideas in order to
try and compare theory with observation. On the other hand, if one
does identify \aD\ with the scale factor, comparison with standard
cosmology becomes more convenient, but this brings in additional
constraints on the underlying inhomogeneous geometry. Thus our
conclusion is that the Buchert formalism is a correct and
tractable averaging scheme, provided all the caveats pointed out
in this paragraph are taken care of. Also, when these caveats have
been taken care of correctly, the Buchert formalism is expected to
give the same physical results as the MG approach. We recall that
in the covariant MG approach also, once a spacetime geometry has
been identified for the averaged manifold \Mbar, a gauge must be
selected for the geometry on the underlying manifold, in order to
explicitly compute the correction scalars for comparison with
observation.

The advantage of the MG approach is that it accomplishes in a neat
package what the Buchert approach, with its attendant caveats,
sets out to do. In the MG approach, there are no unaveraged shear
equations, because the trace of the Einstein equations has been
taken after performing the averaging on the underlying geometry.
Since the averaged geometry is FLRW, the shear is zero by
definition. There is a natural metric on the averaged manifold by
construction, the FLRW metric. The correlations satisfy additional
constraints, given by Eqns. (\ref{correln22}). Thus, once a gauge
has been chosen and if one can overcome the computational
complexity of the averaging operation, the cosmological equations
derived by us in the MG approach are complete and ready for
application, without any further caveats.

In spite of these differences, our equations \eqref{correln26} and
\eqref{correln28} for the volume preserving gauge are strikingly
similar to Buchert's effective FLRW equations and their
integrability condition in the dust case; and in the case of
general $N$, the role of Buchert's dynamical backreaction
$\uD{\bar{\Cal{P}}}$ in \eqns{ep1} and \eqref{incon} is identical
to that of our combination of
$(\tilde{\Cal{P}}^{(2)}+\tilde{\Cal{S}}^{(2)})$ in \eqns{gauge4b}
and \eqref{gauge7}. Concentrating on the volume preserving case,
the structure of the correlation $\Cal{Q}^{(1)}$ is identical to
Buchert's kinematical backreaction $\uD{\Cal{Q}}$ (or
$\uD{\bar{\Cal{Q}}}$ in the general case). The correlation
$\Cal{S}^{(1)}$ appears in place of the averaged $3$-Ricci scalar
$\avgD{\Cal{R}}$ in Buchert's dust equations. This is not
unreasonable since Buchert's $\avgD{\Cal{R}}$ can be thought of as
$\avgD{\Cal{R}} = 6k_\Cal{D}/a_\Cal{D}^2 +\,$corrections, where
$6k_\Cal{D}/a_\Cal{D}^2$ represents the $3$-Ricci scalar on the
averaged manifold which in our case is zero, and hence
$\Cal{S}^{(1)}$ represents the corrections due to averaging.
Further, these similarities are in spite of the fact that our
correlations were defined assuming that a \emph{volume preserving}
gauge averages out to the FLRW $3$-metric in standard form,
whereas Buchert's averaging is most naturally adapted to beginning
with a \emph{synchronous} gauge. This remarkable feature, at least
to our understanding, does not seem to have any deeper meaning --
it simply seems to arise from the structure of the Einstein
equations themselves, together with our assumption
$\rmb{D}_{\bar\Omega}\bZ{a}{b}{i}{j} = 0$. In the absence of this
latter condition, one would have to consider the correlation $3$-
and $4$-forms mentioned earlier, and the structure of the
correlation terms and their ``conservation'' equations would be
far more complicated.

An entirely different outlook towards his approach has been 
emphasized to us by Buchert \cite{buchpriv}. According to Buchert, the  
absence of an averaged manifold \Mbar\ is not to be thought of as a
`caveat', but as a feature deliberately retained `on purpose'. The
actual inhomogeneous Universe is regarded by Buchert as the only
fundamental entity, and the introduction of an averaged Universe is in
fact regarded as an unphysical and unnecessary approximation. As we
mentioned earlier, this is probably the most important difference
between MG and Buchert's approach. In the latter, contact with
observations is to be made by constructing averaged quantities, such
as the scalars defined earlier in this section, and by introducing the
expansion factor $\aD$. The assertion here is that the averaging of
\emph{geometry}, as discussed in MG or in the Renormalization Group
approach of Buchert and Carfora \cite{buch-carf} is not an
indispensable step in comparing the inhomogeneous Universe with actual
observations. The need for averaging of geometry is to be physically
separated from simply looking at effective properties (such as the
constructed scalars) which can be defined for any inhomogeneous
metric. Averaging of geometry becomes relevant if (i) an observer
insists on interpreting the data in a FLRW template model, so that
(s)he needs a mapping from the actual inhomogeneous slice and its
average properties to the corresponding properties in this template,
or (ii) one desires a mock metric, to sort of have a thermodynamic
effective metric to approximate the real one.  (In this context it
should perhaps also be mentioned that the importance of a thin
time-slice approximation of spacetime averaging (as opposed to a
strict spatial averaging) has been stressed also by Buchert
\cite{buchDust}.)

\section{Discussion}
\noindent In this paper we have addressed the issue of
modifications to the standard cosmological equations arising out
of explicitly accounting for the averaging procedure that must
necessarily be performed when studying Cosmology. While this issue
has been dealt with in the literature in various forms, the effect
of such modifications is still a subject of debate, mainly due to
ambiguities in the averaging schemes available. We have applied
the formalism of Zalaletdinov's Macroscopic Gravity (MG) which is
a fully covariant and nonperturbative averaging scheme, in an
attempt to construct gauge independent corrections to the standard
FLRW equations.

We find that one cannot escape the problem of gauge dependence
entirely, which is mainly due to the fact that Weyl's postulate
which is commonly used in the standard approach to Cosmology, must
be refined in the context of averaging in General Relativity, and
one is forced to assume that the FLRW spatial sections in their
natural coordinates arise from averaging the inhomogeneous metric
in a particular gauge. However, the choice of this gauge can at
least formally be left unspecified, and \emph{spacetime scalar}
corrections to the FLRW equations can be constructed. This
partially removes the criticism usually faced by adherents of the
averaging approach, which essentially states that effects of
averaging are likely to be gauge artifacts. We have shown that
these effects are independent of the choice of coordinates in the
averaged manifold, and depend on \emph{one} gauge choice -- a
choice which itself is fundamental to the assumption of large
scale homogeneity and isotropy.

One issue which we did not address was the fact that the
\emph{scale} of averaging is likely to be different at different
epochs. The scale $\Lfrw\sim100 h^{-1}$Mpc which we mentioned in
Section \ref{spatlim}, would be appropriate for the present day
Universe. For the \emph{early} Universe however, this scale will
probably be very different. Nevertheless, since the formalism
simply assumes the existence of a scale at which homogeneity sets
in, without being affected by its actual value, the equations we
have derived within the MG framework will apply to both the
present day as well as to the early Universe, albeit with
different averaging scales. If one assumes that the length scale
of averaging varies slowly compared to the Hubble expansion scale,
then it would be appropriate to apply the formalism independently
to say, Supernovae data and the last scattering epoch. However,
this issue needs to be dealt with more precisely than discussed
here.

This brings us to the most important question to be addressed :
\emph{How} does one apply this formalism without knowing the
``true'' inhomogeneous metric of the Universe? One possible line
of attack may be as follows : In the early Universe at least, it
might be reasonable to assume that the ``true'' metric is a
perturbation around the FLRW metric. Using some standard gauge,
making the necessary assumptions and ensuring that all required
conditions are satisfied, one should then be able to
\emph{explicitly} compute the scalar corrections given the initial
power spectrum of the fluctuations. The behaviour of these
corrections (in the perturbative regime) may yield some insight as
to how the corrections might evolve in the present epoch
(nonlinear regime). One could then construct reasonable models for
the functions $\tilde{\Cal{P}}^{(1)}$, $\tilde{\Cal{S}}^{(1)}$,
$\tilde{\Cal{P}}^{(2)}$ and $\tilde{\Cal{S}}^{(2)}$ defined in
Section \ref{gauge}. Note that in this regime, one needn't worry
about the explicit construction of these quantities -- what one
needs are physically reasonable \emph{models} of the time
evolution of these quantities (see, e.g., \Cite{buchMorphon}). We
hope to commence such a program in the near future.
\label{discuss}

\acknowledgments

\noindent We would like to thank Naresh Dadhich, Friedrich Hehl,
Claus Kiefer, Alok Maharana, Yuri Obukhov, Sarang Sane, Barbara
Sandhofer and Rakesh Tibrewala for useful discussions. It is also
a pleasure to acknowledge insightful comments from Thomas Buchert,
Alan Coley, Syksy R\"as\"anen and Roustam Zalaletdinov on an earlier
version of this paper.

\appendix
\section{}\label{mainapp}
\noindent In this appendix we give proofs of several results that
were used in the text.

\subsection{Analysis of $\rmb{D}_{\bar\Omega}\bar g^{ab}=0$}
\label{app-gbar} \noindent We start with the metric
\begin{equation}
^{(\Cal{M})}ds^2 = g_{00}(t,\rmb{x})dt^2 +
g_{AB}(t,\rmb{x})dx^Adx^B
  \,,
\label{appA1}
\end{equation}
on \Cal{M}\ and assume that it averages out to the FLRW form
(\eqn{spatlim-gauge1}):
\begin{align}
&G_{00} = \avg{\bil{g}_{00}} = -f^2(t) ~~;~~ G_{0A} =
\avg{\bil{g}_{0A}} = 0 ~; \nonumber\\
& G_{AB} = \avg{\bil{g}_{AB}} =
a^2(t)\delta_{AB} \,. \label{appA2}
\end{align}
We will analyze the second relation of \eqn{recap32} and show that
it leads to the result $U^{ij}\equiv\bar g^{ij}-G^{ij}=0$, where
$\barOm{a}{b}$ refers to the connection $1$-forms associated with
$G_{ij}$. For this section we use the notation $H=(1/a)(da/dt)$.
We have
\begin{equation}
\ext\bar g^{ab} + \barOm{a}{j}\bar g^{jb} + \barOm{b}{j}\bar
g^{aj} = 0\,. \label{appA4}
\end{equation}
Consider the three cases ($a=b=0$), ($a=0,b=B$) and ($a=A,b=B$) in
turn. From the first case ($a=b=0$) we can conclude that
\begin{subequations}
\begin{equation}
\bar g^{00}(t,\rmb{x}) = -\frac{k(\rmb{x})}{f^2(t)} \,,
\label{appA7a}
\end{equation}
\begin{equation}
\partial_A k(\rmb{x}) = 2a^2H\delta_{AB}\bar g^{0B} \,.
\label{appA7b}
\end{equation}
\label{appA7}
\end{subequations}
where $k(\rmb{x})$ is a positive definite function (so that the
metric signature is preserved) which arises as an integration
constant and is constrained by \eqn{appA7b}. The second case
($a=0,b=B$) leads to
\begin{subequations}
\begin{equation}
\bar g^{0B} = \frac{m^B(\rmb{x})}{a(t)f(t)}\,, \label{appA9a}
\end{equation}
\begin{equation}
\frac{1}{af}\partial_Jm^B(\rmb{x}) +
\frac{a^2}{f^2}H\delta_{AJ}\bar g^{AB} -
\frac{k(\rmb{x})}{f^2}H\delta{}^B_J = 0\,. \label{appA9b}
\end{equation}
\label{appA9}
\end{subequations}
where $m^B(\rmb{x})$ is a $3$-vector that arises as a constant of
integration like $k(\rmb{x})$, and is constrained by \eqn{appA9b}.
Finally, the last case ($a=A,b=B$) leads to
\begin{subequations}
\begin{equation}
\bar g^{AB} = \frac{1}{a^2(t)}s^{AB}(\rmb{x}) \,, \label{appA11a}
\end{equation}
\begin{equation}
\frac{1}{a^2}\partial_Js^{AB}(\rmb{x}) +
\frac{1}{af}H\left(\delta{}^A_Jm^B(\rmb{x}) +
\delta{}^B_Jm^A(\rmb{x}) \right) = 0\,. \label{appA11b}
\end{equation}
\label{appA11}
\end{subequations}
Here $s^{AB}(\rmb{x})$ is another constant of integration, a
symmetric $3$-tensor. It can be easily argued that since $f$ and
$a$ are not \emph{a priori} related, both sides of \eqn{appA7b}
must vanish, which immediately tells us that the vector
$m^B(\rmb{x})$ must vanish, and the function $k(\rmb{x})$ must be
a constant,
\begin{equation}
k(\rmb{x}) = k = {\rm constant} ~~;~~ m^B(\rmb{x}) = 0\,.
\label{appA12}
\end{equation}
Equations \eqref{appA9b} and \eqref{appA11b} then give us
\begin{equation}
s^{AB}(\rmb{x}) = k \delta^{AB}\,, \label{appA13}
\end{equation}
with the same constant $k$ as in \eqn{appA12}. Finally, putting
everything together we find
\begin{align}
\bar g^{00} = -\frac{k}{f^2} ~~;~~ \bar g^{0A} = 0 ~~&;~~ \bar
g^{AB} =
\frac{k}{a^2}\delta^{AB} \,,\nonumber\\
\Rightarrow \bar g^{ij} = kG^{ij}\,. \label{appA14}
\end{align}
The constant $k$ is not constrained by any of the equations and
appears to be a free parameter in the theory. The modified
Einstein equations \eqref{recap36} show that $k$ can be absorbed
into the averaged energy momentum tensor. We will for simplicity
assume $k$ to be unity thereby obtaining, as required
\begin{equation}
U^{ij} \equiv \bar g^{ij} - G^{ij} = 0\,. \label{appA15}
\end{equation}

\subsection{Analysis of the condition $\avg{\Gamma{}^a_{bc}} =
\,^{(\rm FLRW)}\Gamma{}^a_{bc}$} \label{app-gamma} \noindent Here
we will assume that the line element on \Cal{M}\ is in the volume
preserving gauge
\begin{equation}
^{(\Cal{M})}ds^2=-\frac{d\bt^2}{h(\bt,\rmb{x})} +
  h_{AB}(\bt,\rmb{x})dx^Adx^B\,,
\label{appA16}
\end{equation}
so that the averaging is trivial, and the metric and averages out
to the FLRW line element on \Mbar\ given by
\begin{equation}
^{(\Mbar)}ds^2 = -\frac{d\bt^2}{\avg{h}\!(\bt)} +
  \ba^2(\bt)\delta_{AB}dx^Adx^B \,,
\label{appA17}
\end{equation}
where we used the condition $\avg{1/h} = 1/\avg{h}$ that follows
from $\bar g^{00} = G^{00}$. The conditions $\avg{\Gamma{}^a_{bc}}
= \,^{(\rm FLRW)} \Gamma{}^a_{bc}$ then result in the following :
\begin{subequations}
\begin{align}
&\Gamma{}^0_{00} ~~:~~ \avg{\partial_{\bt}(\ln\sqrt{h})} =
\partial_{\bt}(\ln\sqrt{\avg{h}})\,,
\label{appA18a} \\
&\Gamma{}^0_{0A} ~~:~~ \avg{\partial_A(\ln\sqrt{h})} = 0\,,
\label{appA18b} \\
&\Gamma{}^A_{00} ~~:~~
\avg{\frac{h^{AB}}{h}\partial_B(\ln\sqrt{h})} = 0\,,
\label{appA18c} \\
&\Gamma{}^A_{0B} ~~:~~ \avg{\frac{1}{\sqrt h}\Theta{}^A_B} =
H\delta{}^A_B \,,
\label{appA18d} \\
&\Gamma{}^0_{AB} ~~:~~ \avg{\sqrt{h}\Theta_{AB}} =
\avg{h}\ba^2H\delta_{AB} \,,
\label{appA18e} \\
&\Gamma{}^A_{BC} ~~:~~ \avg{\,^{(3)}\Gamma{}^A_{BC}} = 0 \,.
\label{appA18f}
\end{align}
\label{appA18}
\end{subequations}
\eqns{appA18b} and \eqref{appA18f} are consistent with each other
since $\,^{(3)}\Gamma{}^A_{BA} = \partial_B(\ln\sqrt{h})$, and
\eqn{appA18c} is consistent with the assumption \eqn{recap31a}.
The trace of \eqn{appA18d} gives $\avg{(1/\sqrt h)\Theta} = 3H$.
However, using \eqn{correln2} we have $(1/\sqrt h)\Theta = N\Theta
=
\partial_{\bt}(\ln\sqrt{h})$, and combined with \eqn{appA18a} this
gives
\begin{equation}
\frac{1}{2}\partial_{\bt}(\ln\avg{h}) = 3\partial_{\bt}(\ln\ba)
\Rightarrow \avg{h} = \ba^6\,, \label{appA19}
\end{equation}
where we have set an arbitrary proportionality constant
(representing rescaling of the time coordinate by a constant) to
unity. This establishes the last equality in \eqn{spatlim10}.

Finally, consider the trace
$(\avg{h^{AB}}/\avg{h})\avg{\sqrt{h}\Theta_{AB}}$ : using the
condition $\bar g^{AB} = G^{AB}$, \eqn{appA18e} and the trace of
\eqn{appA18d}, this gives us
\begin{align}
\frac{\avg{h^{AB}}}{\avg{h}}\avg{\sqrt{h}\Theta_{AB}} &=
\frac{1}{\avg{h}}\frac{\delta^{AB}}{\ba^2}
\avg{\sqrt{h}\Theta_{AB}} \nonumber\\
&= 3H = \avg{\frac{1}{\sqrt h}\Theta} =
\avg{\frac{h^{AB}}{h}(\sqrt{h}\Theta_{AB})}\,. \label{appA20}
\end{align}
On using the condition \eqref{recap31a} this leads to
\begin{equation}
\left(\frac{\avg{h^{AB}}}{\avg{h}} -
\avg{\frac{h^{AB}}{h}}\right)\avg{\Gamma{}^0_{AB}} = 0\,,
\label{appA21}
\end{equation}
which is consistent with the assumption
\begin{equation}
\frac{\avg{h^{AB}}}{\avg{h}} = \avg{\frac{h^{AB}}{h}}\,.
\label{appA22}
\end{equation}

\label{refer}


\begin{thebibliography}{100}
\bibitem{note1}
We are making the assumption $N^A=0$ in the volume preserving
gauge for algebraic convenience only -- in our case this
assumption cannot be justified by the absence of vorticity. A more
detailed (and complicated) analysis should retain an arbitrary
$N^A$ in the inhomogeneous geometry, and make assumptions about
its average -- such as $\avg{N_A}=0$ for example.
\bibitem{note2}
In fact, the shape of the \emph{spacetime} averaging region can
also be left arbitrary, provided the spatial averaging limit is
correctly taken as the limit in which one is essentially
integrating over a spatial domain only. We will not venture into
this complication.
\end{thebibliography}
\end{document}